\documentclass[12pt, letterpaper, onecolumn]{article}
\pdfoutput=1 

\usepackage[utf8]{inputenc}
\usepackage[dvipsnames]{xcolor}

\usepackage[top=1in, bottom=1in, left=1in, right=1in]{geometry}

\usepackage{graphicx}
\usepackage[scaled]{helvet}
\usepackage{boldmath}
\usepackage{amsmath, amsthm, amssymb}
\usepackage{bm}
\usepackage{array, booktabs, multirow, tabularx}
\usepackage{setspace}
\usepackage{xspace}
\usepackage{ulem}
\usepackage{hyperref}

\usepackage{color}
\usepackage{xcolor}
\usepackage[toc,acronym,section=section,nonumberlist,style=super3col]{glossaries}
\usepackage[nooneline,labelsep=period]{caption}
\usepackage[]{float}

\DeclareMathOperator*{\argmin}{argmin}

\glsdisablehyper
\newacronym{tr}{TR}{repetition time}
\newacronym{snr}{SNR}{signal-to-noise ratio}
\newacronym{ssEPI}{ssEPI}{single-shot echo-planar
imaging}
\newacronym[\glslongpluralkey={convolutional neural networks}, \glsshortpluralkey={CNNs}]{cnn}{CNN}{convolutional neural network}

\newcommand{\bs}{\boldsymbol}

\begin{document}

	\date{}
	
	\title{\Large{\bf Robust partial Fourier reconstruction for diffusion-weighted imaging using a recurrent convolutional neural network}}
	\author{Fasil Gadjimuradov$^{1,2}$, Thomas Benkert$^{2}$, Marcel Dominik Nickel$^{2}$, \\Andreas Maier$^{1}$}
	
	\maketitle
	\thispagestyle{empty}
	
	\begin{itemize}
	\item[1] Pattern Recognition Lab, Department of Computer Science, Friedrich-Alexander University Erlangen-Nürnberg, Erlangen, Germany
	\item[2] Magnetic Resonance Applications Predevelopment, Siemens Healthcare GmbH, Erlangen, Germany
	\end{itemize}
	
	\vfill	
	
	\noindent \begin{tabular*}{\textwidth}{ll}
		Running Head: & Robust PF reconstruction for DWI using a recurrent CNN\\
		Correspondence to: &Fasil Gadjimuradov\\
		&Pattern Recognition Lab\\ 
		&Department of Computer Science\\
		&Friedrich-Alexander University Erlangen-Nürnberg\\
		&Martensstr. 3, D-91058 Erlangen, Germany\\
		&E-mail: fasil.gadjimuradov@fau.de
	\end{tabular*} \vspace{.5cm}
	
	\noindent Preliminary data of this work was presented at the 29th Annual Meeting of ISMRM, 2021. \vspace{.5cm}

	\noindent
	Number of words (abstract): 247\\
	Number of words (body): approx. 5,600\\
	Number of figures: 8\\
	Number of tables: 1\\
	Number of references: 35\vspace{1cm}
	
	\noindent
	Typeset version published in Magnetic Resonance in Medicine is available under: \\
	\url{https://onlinelibrary.wiley.com/doi/10.1002/mrm.29100}

	\linespread{1.5}	
	
	\clearpage	
	\section*{Abstract}
	\small
	\vspace{0.3cm}
	\noindent
	\textit{\textbf{Purpose:}} To develop an algorithm for robust partial Fourier (PF) reconstruction applicable to diffusion-weighted (DW) images with non-smooth phase variations.
		
	\noindent
	\textit{\textbf{Methods:}} Based on an unrolled proximal splitting algorithm, a neural network architecture is derived which alternates between data consistency operations and regularization implemented by recurrent convolutions. In order to exploit correlations, multiple repetitions of the same slice are jointly reconstructed under consideration of permutation-equivariance. The algorithm is trained on DW liver data of 60 volunteers and evaluated on retrospectively and prospectively sub-sampled data of different anatomies and resolutions.
	
	\noindent
	\textit{\textbf{Results:}} The proposed method is able to significantly outperform conventional PF techniques on retrospectively sub-sampled data in terms of quantitative measures as well as perceptual image quality. In this context, joint reconstruction of repetitions as well as the particular type of recurrent network unrolling are found to be beneficial with respect to reconstruction quality. On prospectively PF-sampled data, the proposed method enables DW imaging with higher signal without sacrificing image resolution or introducing additional artifacts. Alternatively, it can be used to counter the TE increase in acquisitions with higher resolution. Further, generalizability can be shown to prospective brain data exhibiting anatomies and contrasts not present in the training set.
	
	\noindent
	\textit{\textbf{Conclusion:}} This work demonstrates that robust PF reconstruction of DW data is feasible even at strong PF factors in anatomies prone to phase variations. Since the proposed method does not rely on smoothness priors of the phase but uses learned recurrent convolutions instead, artifacts of conventional PF methods can be avoided. \vspace{1cm}
		
	\noindent 	
	\textit{\textbf{Key words}}:
		Diffusion-weighted Imaging;
		Deep Learning;
		Partial Fourier Imaging;
		Liver imaging;
		Learning-based Reconstruction
	
	\clearpage
	\section{Introduction} \label{sec:introduction}
	Due to its ability to visualize regions of restricted diffusion, diffusion-weighted imaging (DWI) has been shown to be valuable in detecting and assessing lesions in several clinical applications \cite{LiverDWI, LIRADS, PIRADS, brainDWI}. However, DWI is an inherently signal-starved imaging technique since the diffusion encoding requires the application of strong magnetic gradients which dephase parts of the spins and reduce total net magnetization. In addition, DWI often employs \gls{ssEPI} -- which has lower scan times than its multi-shot counterpart and does not have to compensate for inter-shot motion -- but suffers from T2*-decay for long echo-trains. In order to alleviate T2*-related effects and increase \gls{snr}, partial Fourier (PF) sampling along the phase-encoding direction can be employed which reduces the distance to the $k$-space center from one side and, hence, results in shorter echo time (TE). The ratio between the acquired portion of $k$-space and the total $k$-space is referred to as PF factor (PFF), with typical values of 5/8, 6/8 or 7/8.

	Reconstructing an image from an asymmetrically sampled $k$-space is a non-trivial task given that the image at hand is not real-valued. The central idea of PF reconstruction is based on the observation that the phase of MR images usually has relatively low spatial resolution and, hence, can be estimated from the partially acquired data. Conventional PF techniques, such as Homodyne \cite{Homodyne} and Projection Onto Convex Sets (POCS) \cite{POCS}, make explicit assumptions on the smoothness of the phase. Both methods perform phase correction using a low-resolution phase estimate which is computed from the symmetrically sampled portion of the data. While Homodyne is a one-step approach which involves ramp filtering in frequency domain, POCS is an iterative method that alternates between phase correction and enforcing data consistency. 
	
	 Although smoothness priors are appropriate in many MR applications, they may be unsuitable in the context of DWI which oftentimes exhibits rapid phase variations, especially in motion-prone anatomies, such as the abdomen (see Supporting Information Figure \ref{sfig:phase_liver}). In the case of strong PFFs which reduce the range of symmetrically sampled data in $k$-space, conventional methods oftentimes introduce artifacts because the respective smooth phase estimate cannot appropriately represent the true image phase. Furthermore, phase variations may become strong enough to shift the $k$-space center out of the partially sampled region resulting in severe, irreversible signal loss \cite{Storey}.
	
	In recent years, Deep Learning-based approaches found their way into the field of under-sampled MR reconstruction, including PF reconstruction \cite{PF1, PF2, PF3, Muckley, DRPF}. Using the expressive power of \glspl{cnn}, these methods try to learn a reconstruction procedure from large sets of training data and thereby circumvent weaknesses of hand-crafted image priors. Typically, training is performed in a supervised manner by acquiring ground-truth data which is retrospectively sub-sampled to create inputs of the network. The parameters of the \gls{cnn} are then tuned towards the desired behavior by backpropagating gradients from a loss function that compares the network output with the ground-truth. In \cite{Muckley}, it was shown that a \gls{cnn} trained on PF reconstruction of natural images from the ImageNet database \cite{ImageNet} was able to generalize to DW brain images during test time. However, the fact that ImageNet data is real-valued, which necessitated the simulation of random phases, constitutes a limiting factor. Further, in the context of medical image reconstruction, one general drawback of the networks which were proposed in \cite{PF1, PF2, Muckley} is that they perform direct image-to-image mappings and consequently lack assurance of data consistency. Unrolled networks that attempt to emulate iterative optimization schemes are known to perform better as they implement a gradual refinement of estimates and encode knowledge of the physics of the acquisition \cite{PL, Known, Modl}, e.g.\ by incorporating the forward model in a data consistency step. This makes deep neural networks more interpretable as well and can therefore increase acceptance among clinicians. The method proposed in \cite{DRPF} was derived by unrolling a relaxed version of the POCS algorithm but replacing the phase correction step by a recurrent CNN.
	
	This work aims to further develop the preliminary approach presented in \cite{DRPF} by making more efficient use of correlations among multiple realizations of the same slice and eliminating the dependency on their ordering. In addition, training and evaluation are performed on liver DWI which is known to be very challenging in the context of PF reconstruction due to strong, motion-induced phase variations. The proposed method is evaluated quantitatively as well as qualitatively by comparison with conventional PF techniques and other Deep Learning-based approaches. A reconstruction method that works robustly even at strong PFFs in anatomies with high-frequency phase components facilitates the applicability of PF to DWI. Provided that image sharpness is restored reliably, the SNR gain due to PF-induced TE reduction can increase the diagnostic value of DWI. Alternatively, the increased SNR can be traded against scan time by reducing the number of acquired repetitions or used to compensate for the increase in TE related to acquisitions with higher spatial resolutions or motion-compensated diffusion preparation \cite{Karampinos, MODI}.

	\section{Theory} \label{sec:theory}
	
	\subsection{Inverse reconstruction problem}
	Assuming single-channel data $\bs{y} \in \mathbb{C}^M$, the forward model in PF-sampled MRI can be expressed as
\begin{equation}
	\bs{y} = \bs{Ax} + \bs{n}
\end{equation}
where $\bs{x} \in \mathbb{C}^N$ is the complex-valued image, $\bs{A} \in \mathbb{C}^{M \times N}$ is the linear forward operator, encoding both PF-sampling and discrete Fourier transform, and ${\bs{n} \in \mathbb{C}^M}$ represents complex Gaussian noise. Due to non-injectivity associated with the sub-sampling ($M < N$) and due to noise, recovering $\bs{x}$ from its measurements $\bs{y}$ is an ill-posed inverse problem. It can be addressed by minimizing a regularized objective function of the form
\begin{equation}
\label{Regularized}
	\argmin_{\bs{x}} \mathcal{R}(\bs{x}) + \nu \, \mathcal{D}(\bs{Ax}, \bs{y})
\end{equation}
where $\mathcal{R}$ and $\mathcal{D}$ represent regularization and data terms, respectively. Both terms are weighted against each other by some parameter $\nu > 0$. Using $\mathcal{D}(\bs{Ax}, \bs{y}) = \frac{1}{2}||\bs{Ax}-\bs{y}||^2_2$ which is the natural choice for normally distributed noise, a proximal splitting algorithm can be employed to approximate the solution of Equation \eqref{Regularized}. Given iteration index $k \in \mathbb{N}$, the proximal operators are computed on both terms in an alternating fashion with step size $\mu_k \geq 0$:
\begin{subequations}
\begin{alignat}{3}
	\bs{z}_k &= \text{prox}_{\mu_k, \mathcal{R}}(\bs{x}_{k-1}) &&= \argmin_{\bs{z}} \mathcal{R}(\bs{z}) + \frac{\mu_k}{2} ||\bs{x}_{k-1} - \bs{z}||^2_2  \label{proxR}\\
	\bs{x}_k &= \text{prox}_{\mu_k, \nu \mathcal{D}}(\bs{z}_k) &&= \bs{z}_k - \frac{1}{1+\lambda_k}\bs{A}^*(\bs{A}\bs{z}_k-\bs{y}) \label{proxD}
\end{alignat}
\end{subequations}
where $\bs{x}_0 = \bs{A}^* \bs{y}$ and $\lambda_k = \frac{\mu_k}{\nu}$. 

Since proximal mappings constitute generalizations of projections, it can be easily seen, that POCS is in fact a realization of this scheme for $\lambda_k = 0$ and $\mathcal{R} = \iota_{\Phi}$, where $\iota_{\Phi}$ is the indicator function of the set of images consistent with the low-resolution phase estimate $\bs{\phi} \in \mathbb{R}^N$:
\begin{equation*}
	\iota_{\Phi}(\bs{x}) = 
	\begin{cases}
		0, & \text{if}\ \angle \bs{x} = \bs{\phi}\\
		\infty, & \text{else}
	\end{cases} \quad \>.
\end{equation*}

	\subsection{Unrolled recurrent network architecture}
	The proposed reconstruction method is derived from unrolling the aforementioned iterative scheme for a fixed number of iterations $K$ and parametrizing the proximal mapping onto the regularizer term in Equation \eqref{proxR} by a \gls{cnn} instead of relying on hand-crafted priors. 

	In general, there are two different ways of employing a \gls{cnn} for the regularization step in unrolled frameworks: 1) weight-sharing and 2) cascading, i.e.\ parametrizing distinct network instances for iterations as in \cite{Schlemper, VN}, for example. The first category includes ``plug-and-play" approaches in which a \gls{cnn} is trained in isolation and applied in an iterative scheme afterwards \cite{Meinhardt}. While the first strategy has the advantages of requiring less parameters and not necessarily being tied to a fixed number of iterations, the latter offers more flexibility in the set of performed operations over iterations, however, at the cost of an increased risk of overfitting. In order to effectively combine the benefits of both strategies, recurrent architectures can be employed which implement weight-sharing but simultaneously offer operational flexibility as the network's behavior might be slightly modified based on the results of earlier iterations.

\begin{figure}[tbp]
	\centering
	\includegraphics[width=\textwidth]{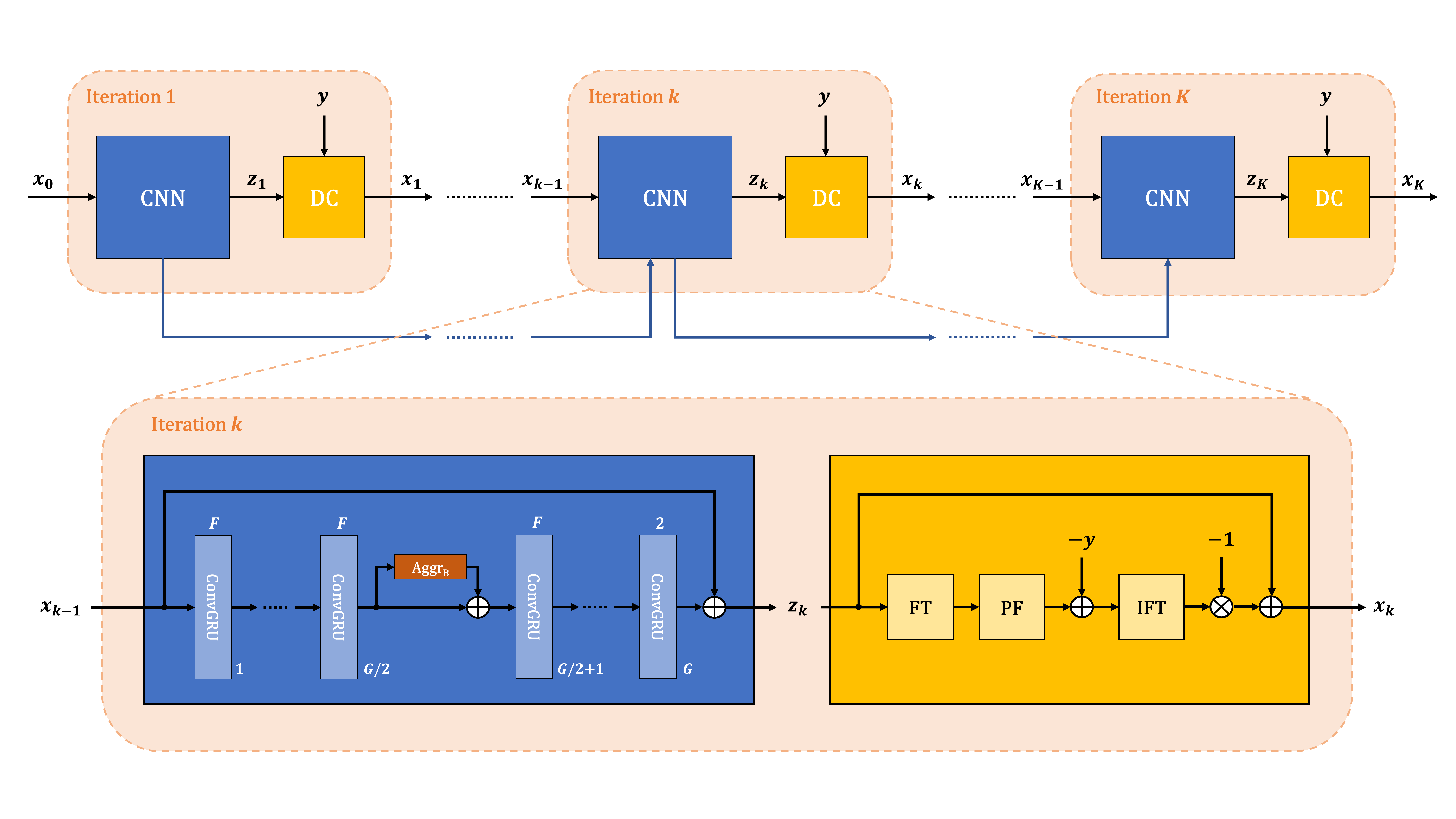}
	\caption{Overview of the employed network architecture which is unrolled for $K$ iterations. Within every iteration it alternates between a CNN-based regularization and a data consistency block (DC). The latter consists of known operators only, such as (inverse) Fourier transform ((I)FT) and PF-sampling (PF). The regularization network is implemented by a stack of $G$ convolutional gated recurrent units (ConvGRUs) which propagate an internal memory state along iterations as visualized by the blue lines. Apart from the last unit, every unit uses $F$ feature maps for convolutions. After $G/2$ of the units, a batch aggregation (Aggr$_{\text{B}}$) is performed which allows to distribute shared information across the batch.}
	\label{fig:architecture}
\end{figure}

	The resulting unrolled network architecture is shown in Figure \ref{fig:architecture}. Here, recurrence is realized by using a stack of $G$ convolutional gated recurrent units (ConvGRUs) \cite{GRU, Ballas} which keep track of an internal memory state across iterations. Within each iteration, the CNN-based regularization is followed by a data consistency block implemented according to Equation \ref{proxD}. The architecture is similar to the one used in \cite{DRPF} but differs with respect to the implementation of joint reconstruction of image repetitions as pointed out in the following.

	\subsection{Joint processing of image sets}
	The fact that DWI has inherently low SNR necessitates the acquisition of multiple repetitions, implying potential benefits of joint reconstruction. In \cite{DRPF}, this was attempted by applying 3-D convolutions to repetitions arranged in a 3-D stack. When operating on sets of images, however, two central requirements need to be fulfilled: 1) the ability to handle different set sizes and 2) permutation-equivariance, i.e.\ the result should not depend on the ordering of the set elements. While the first requirement can be satisfied when following the 3-D stacking strategy, it can be easily seen that permutation-equivariance cannot be ensured as a modified order of the images in the stack will modify the network output.
	
	Therefore, the concept of \textit{Deep Sets} \cite{DeepSets} was employed in this work. Here, the set of available repetitions is interpreted as a typical input batch. In order to still exploit information shared among the repetitions, permutation-invariant pooling operations (such as mean or maximum computations) across the batch dimension are performed within the network. Besides the fact that the aforementioned requirements are fulfilled, this concept also allows to employ 2-D convolutions which are more efficient with respect to memory and run-time than 3-D convolutions.

	\section{Methods} \label{sec:methods}
	
	\subsection{Data}
	For training and evaluation purposes, 68 volunteers were scanned on 1.5 and \mbox{3\,T} MR scanners (MAGNETOM, Siemens Healthcare, Erlangen, Germany) after obtaining written informed consent. While the validation and test sets comprised the data of four volunteers each, the remaining 60 data sets were used for training. Using a prototypical free-breathing \gls{ssEPI} sequence, liver DWI was acquired with one diffusion direction (diagonal mode), two $b$-values (50 and \mbox{800\,s/mm$^2$}) and phase-encoding along the anterior-posterior (AP) direction. Further, the following acquisition parameters were used: \mbox{$\text{TE} = [52, 61]$\,ms}, \mbox{$\text{TR} = [5500, 6300]$\,ms}, \mbox{$\text{matrix size} = 134 \times 108$}, isotropic in-plane \mbox{$\text{resolution} = [2.8, 3.2]$\,mm$^2$}, and slice \mbox{$\text{thickness} = 5$\,mm}, where values that varied across measurements are reported as intervals. Given that 30 to 35 axial slices were collected per volunteer, training, validation and test sets comprised 1960, 130 and 130 image slices, respectively. In order to augment the data set prospectively, 15 and 60 repetitions were collected per slice for the low and high $b$-value, respectively, which corresponds to three times the number used in typical clinical protocols. Data was acquired without PF-sampling and with a parallel acceleration factor of 2 due to echo-train length limitations. Ground-truth data was obtained by performing a conventional in-line GRAPPA reconstruction \cite{GRAPPA} followed by coil combination.
	
	Network inputs were generated by retrospective PF-sampling of the coil-combined data followed by an inverse Fourier transform. The reasoning behind that was to ensure modularity of the proposed method. Given that parallel acceleration is a standard setting in ssEPI sequences in order to avoid negative effects of excessively long echo-trains, this approach would allow to employ any parallel reconstruction method beforehand. Also, if the goal was to perform both parallel and PF reconstruction in a single framework, it would have to be tailored to a particular combination of parallel acceleration and PF factor which would require training of multiple network instances for different scenarios. Further, since conventional PF methods are typically applied to coil-combined data following some kind of parallel reconstruction, comparison with those techniques is facilitated as differences in performance would not originate from differences with respect to parallel reconstruction/de-aliasing.
	
	\subsection{Training and implementational details}
	\label{Training}
	Complex-valued images were normalized by the 98th percentile value of the corresponding magnitude images and were flipped along the readout direction with a probability of 50\,\% for the sake of data augmentation during training. An input batch was obtained by taking a random subset comprising one third of the available repetitions for a given slice, such that the network was provided with varying realizations over training epochs. Real and imaginary parts were interpreted as separate channels.

	Given the batch of $B$ output repetitions after $K$ iterations \mbox{$\{\bs{x}_K^1, ..., \bs{x}_K^B\}$}, the magnitude average across the batch was obtained according to \mbox{$\bs{\overline{x}}_{K} = \frac{1}{B} \sum_{b=1}^B |\bs{x}_K^b|$} where $|\cdot|$ is the element-wise magnitude operation. Using the average of the ground-truth repetitions $\bs{\overline{x}}_{\text{GT}}$ computed in the same manner, end-to-end training of the network was performed by minimizing a loss function $\mathcal{L}$ of the form
\begin{equation}
	\mathcal{L} = \mathcal{L}_1(\bs{\overline{x}}_{K}, \bs{\overline{x}}_{\text{GT}}) + 0.5 \cdot \mathcal{L}_{\text{perc}}(\bs{\overline{x}}_{K}, \bs{\overline{x}}_{\text{GT}})
\end{equation}
where $\mathcal{L}_1$ represents the L$_1$-penalized difference averaged over all pixels and $\mathcal{L}_{\text{perc}}$ is a perceptual loss metric introduced in \cite{lpips}. Given that the goal was to restore image resolution and per-pixel metrics such as L$_1$ do not penalize image blur heavily enough, the perceptual loss which calculated normalized distances in a higher-level feature space of a pre-trained neural network aided in promoting sharp reconstructions. Since the focus was still on ensuring high pixel fidelity, the first loss term was weighted twice as strong. Note that no phase information contributed to the loss function. Network parameters were initialized using the method presented in \cite{He} and were optimized for 200 epochs using Adam \cite{Adam} with a learning rate of $5 \cdot 10^{-4}$ (\mbox{$\beta_1 = 0.9$}, \mbox{$\beta_2 = 0.999$}).

	The proposed architecture was unrolled for $K = 5$ iterations, while $G = 10$ ConvGRUs were used within the regularizing CNN. Except for the last unit which was constrained to have 2 feature channels to accommodate for the complex nature of the output data, ConvGRUs used $F = 32$ feature channels. All convolutions used kernels of size $3\!\times\!3$ with zero-padding, such that the image size was kept constant throughout. Exploitation of shared information among repetitions was implemented by maximum aggregation after half of the ConvGRUs (see Figure \ref{fig:architecture}). Given the batch of $B$ feature maps $\{\bs{f}^1, ..., \bs{f}^B\}$, the aggregated feature map was simply computed as the channel- and pixel-wise maximum over the batch: $\bs{f}_{\text{aggr}} = \text{max}(\{\bs{f}^1, ..., \bs{f}^B\})$. In the case of mean aggregation, which was used for evaluations as well, the aggregated feature map would be obtained as $\bs{f}_{\text{aggr}} = \frac{1}{B} \sum_{b=1}^B \bs{f}^b$, accordingly. In both cases, permutation-equivariance was ensured by updating the feature maps according to: $\bs{\tilde{f}}^b = \bs{f}^b + \bs{f}_{\text{aggr}}$ for $b=1,..,B$.

Analogous to POCS, hard projections were performed for data consistency by setting $\lambda_k = 0$ in Equation \eqref{proxD} for all iterations. Thereby, it is ensured that only the non-acquired parts of the $k$-space are estimated. Note that $\lambda_k$ alternatively can be chosen to be greater than 0 in order to allow deviations from the measured data which is required for reconstructions that involve denoising. For example in \cite{DRPF}, $\lambda_k$ were implemented as learnable parameters. However, in this work denoising was not desired in order to ensure comparability with conventional techniques which do not include denoising either.

	\subsection{Evaluation}		
	Qualitative evaluations of representative reconstructions were conducted for the proposed method which will be referred to as \textit{Deep Recurrent Partial Fourier} (DRPF) in the following. For cases in which ground-truth data was available, quantitative measures in the form of peak signal-to-noise ratio (PSNR) and structural similarity (SSIM) were reported as well. Since conventional methods perform sufficiently well in the case of a PFF of 7/8, only factors of 5/8 and 6/8 were considered during evaluations. Concerning the proposed method, distinct network instances were trained and evaluated for each PFF separately. 
	
	\subsubsection{Retrospectively sub-sampled liver DWI}
	Using the data from the test set, the reconstructions generated by DRPF were validated by qualitative and quantitative comparisons with ground-truth images. POCS was included into the comparison because it is a commonly used state-of-the-art method in clinical practice. POCS was performed for 5 iterations, although approximate convergence could generally be observed within as few as 3 iterations.

	\subsubsection{Prospectively sub-sampled liver DWI}
	\label{sec:prospect_data}
	In order to validate the reconstruction quality of the proposed method on prospectively PF-sampled data, three additional data sets were acquired using a PFF of 5/8. Further, the effect of PF-induced TE reduction was demonstrated by comparing the data of a selected volunteer from the test set with two additionally acquired scans of the same volunteer with higher spatial resolution: one without PF-sampling and one with prospective PF-sampling of 5/8. Compared to the data set with standard resolution, the matrix size was increased from \mbox{$134 \times 108$} to \mbox{$180 \times 146$} while keeping the FOV constant, corresponding to an isotropic resolution increase from \mbox{2.8\,mm} to \mbox{2.1\,mm} for both acquisitions. Keeping TR fixed, TE was chosen as minimal as possible in all scans. A detailed comparison of relevant acquisition parameters is provided in Supporting Information Table \ref{stab:acqparam_prospect}.
	
	\subsubsection{Generalization to brain DWI}
	An important aspect with respect to learning-based methods is the ability to generalize to data that is distinct from instances seen during training. The robustness of the proposed method was demonstrated by applying it to a case which deviated from the training set in terms of anatomy, contrast and resolution. For this purpose, brain data was acquired using a PFF of 5/8 with (\mbox{$b=1000$\:s/mm$^2$}, 4 diffusion directions, 4 repetitions) and without diffusion-weighting (\mbox{$b=0$\:s/mm$^2$}, 3 repetitions). The remaining acquisition parameters are listed in Supporting Information Table \ref{stab:acqparam_brain}.
	
	\subsubsection{Benefits of joint reconstruction}
	In order to quantify the benefits of jointly reconstructing multiple repetitions in the context of DWI, the proposed architecture, which used maximum aggregation over the set of repetitions, was compared against two other realizations which used mean aggregation and no aggregation at all, respectively, meaning that correlations across repetitions were not accessible in the latter case.	
	
	\subsubsection{Comparison of unrolling strategies}
	The approach of unrolling a network using a recurrent CNN was compared against two other unrolling strategies which used simple weight-sharing across iterations and a cascade of CNNs, respectively. Both employed a ResNet architecture \cite{ResNet} which approximately matched the number of parameters of the proposed recurrent architecture and used maximum aggregation after half of the residual blocks (see Supporting Information Figure \ref{sfig:resnet}). A base regularizer network (one iteration with data consistency) was pre-trained for 50 epochs with a learning rate of $5 \cdot 10^{-4}$ and then used to initialize both the weight-sharing and cascading architectures, where a distinct copy was used for each iteration in the latter case. Both architectures were then trained for another 150 epochs with a learning rate of $1 \cdot 10^{-4}$. This training strategy showed better convergence and final performance than direct end-to-end training in both cases. The remaining training parameters were the same as described in \ref{Training}.

	\section{Results} \label{sec:results}
	
	\subsection{Retrospectively sub-sampled liver DWI}
\begin{figure}[tbp]
	\centering
	\includegraphics[width=.9\textwidth]{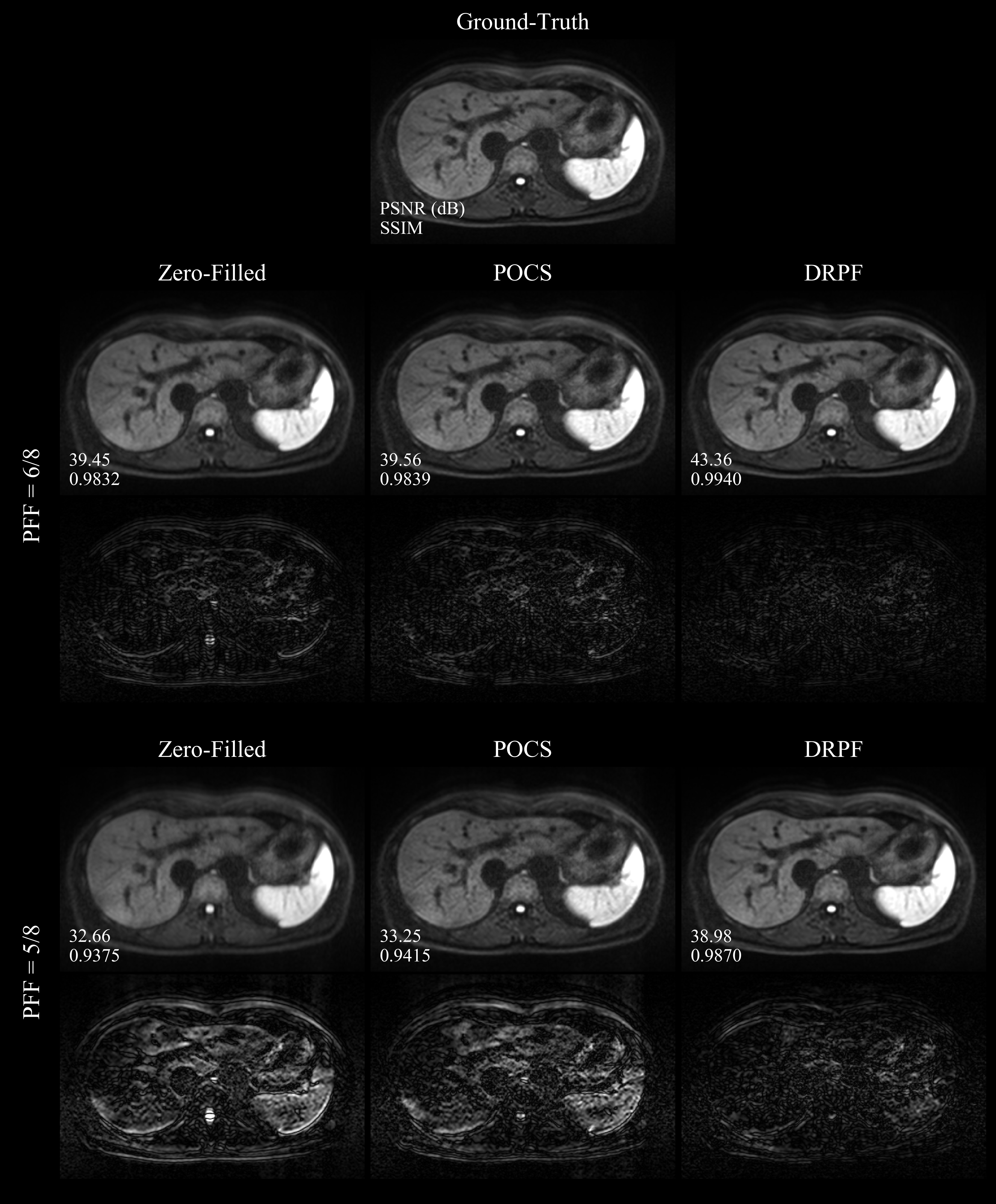}
	\caption{Reconstruction quality on a representative retrospectively PF-sampled liver slice (magnitude average of 20 repetitions acquired at \mbox{1.5\,T} and a $b$-value of \mbox{800\,s/mm$^2$}). Top row: ground-truth image. Second and third row: reconstructions and corresponding difference images (magnified by a factor of 5) produced by zero-filling, POCS and DRPF at a PFF of 6/8. Fourth and fifth row: same as above for a PFF of 5/8.}
	\label{fig:qual_retro}
\end{figure}

	Figure \ref{fig:qual_retro} shows representative reconstruction results for zero-filling, POCS and DRPF on retrospectively PF-sampled liver data acquired at \mbox{$b=800$\,s/mm$^2$}. At a PFF of 6/8, zero-filling introduces mild, high-frequency blurring along AP. Both POCS and DRPF are able to alleviate it, however, the POCS reconstruction exhibits more noise and slightly more residual blurring. Differences between the reconstructions become more apparent for a PFF of 5/8. Here, the sub-sampling results in more severe blurring but also leads to signal voids in some image regions, such as the left hepatic lobe and the spleen. POCS is not able to recover the high-frequency components associated with the signal loss and causes significant noise amplification. In contrast, DRPF produces a sharp image with restored liver signal while not introducing any additional noise. A qualitative evaluation of the corresponding images acquired at a $b$-value of \mbox{50\,s/mm$^2$} is provided in Supporting Information Figure \ref{sfig:qual_retro_b50}.

\begin{figure}[tbp]
	\centering
	\includegraphics[width=\textwidth]{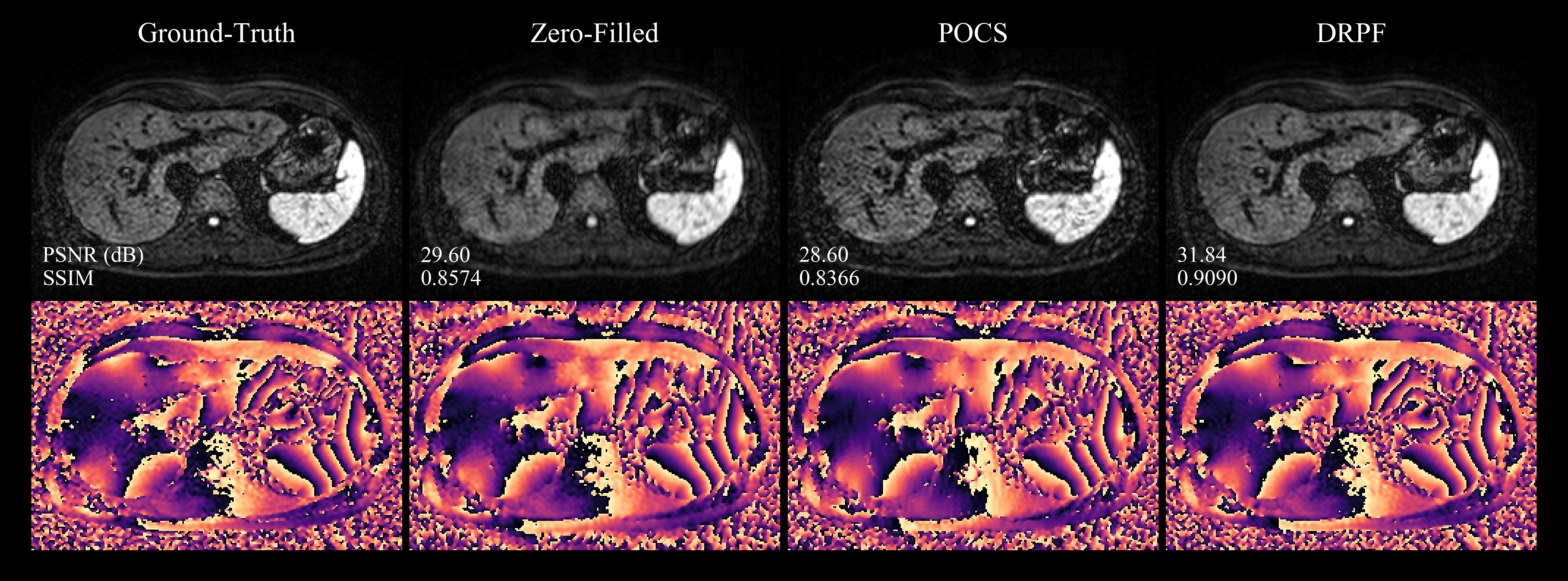}
	\caption{Reconstruction quality of zero-filling, POCS and DRPF on a retrospectively PF-sampled (\mbox{$\text{PFF} = 5/8$}) single repetition from the data used in Figure \ref{fig:qual_retro}. First row: magnitude images. Second row: phase images.}
	\label{fig:qual_retro_single}
\end{figure}

	In order to further examine the effects of the different reconstruction methods, evaluations on a single repetition were conducted which allowed to visualize phase reconstructions as well. As shown in Figure \ref{fig:qual_retro_single}, the aforementioned observations hold true for the magnitude images of the single repetitions as well, i.e.\ POCS is not able to recover regions affected by signal loss and introduces additional noise. Note that the artifacts appear more severe in the single repetitions since they tend to average out in the combined images. The location of the signal void in the left liver lobe coincides with high-frequency structures in the corresponding phase image which cannot be reconstructed properly by POCS. In fact, the phase produced by POCS has even lower resolution than the phase of the zero-filled image since the former only uses the symmetrically sampled portion of the $k$-space. On the other hand, the proposed method is able to accurately reconstruct high-frequency structures in both magnitude and phase images.

\begin{figure}[tbp]
	\centering
	\includegraphics[width=\textwidth]{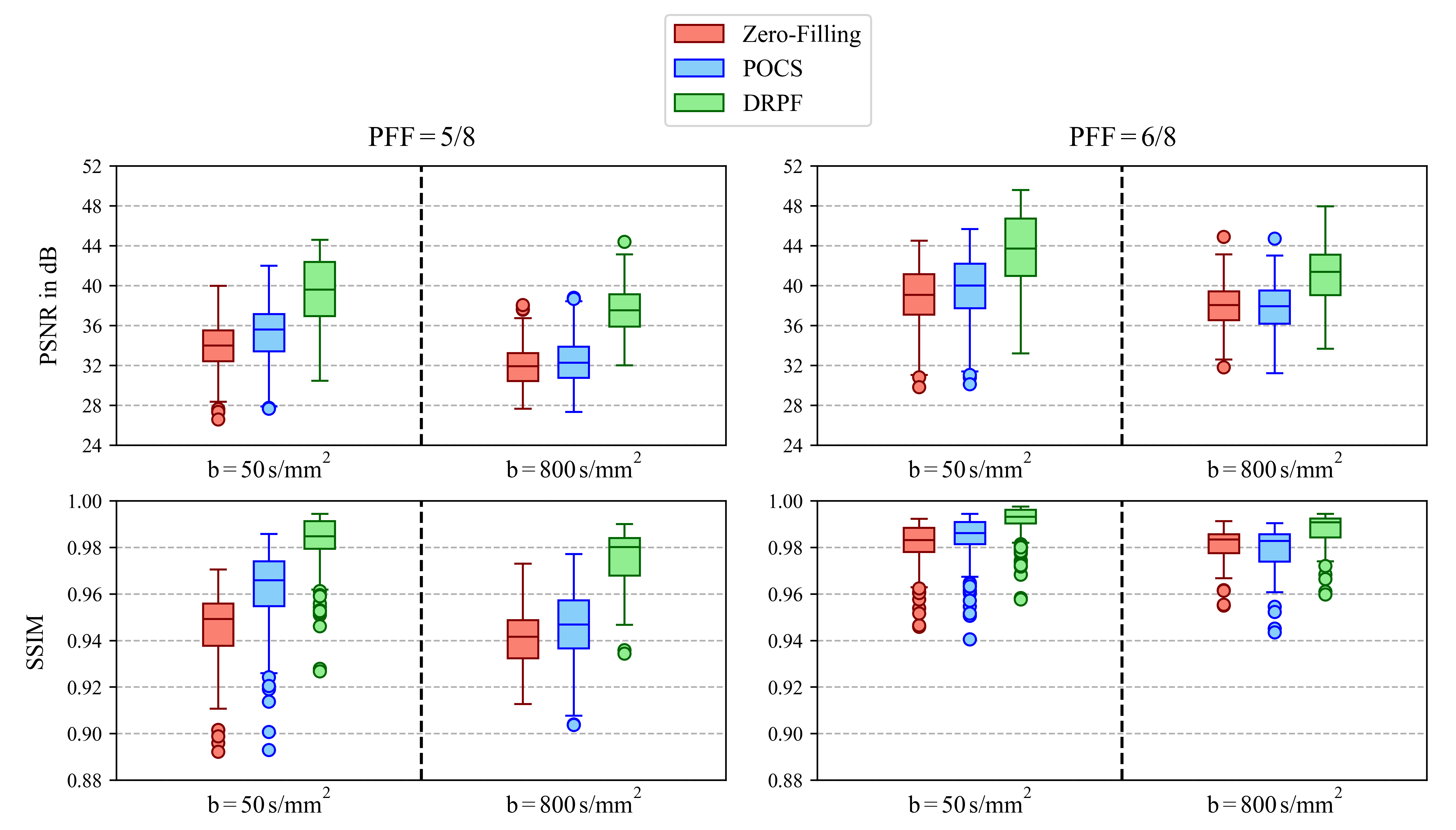}
	\caption{Quantitative evaluation of zero-filling, POCS and DRPF on the retrospectively PF-sampled test set with respect to two metrics (PSNR: first row, SSIM: second row) as well as two PFFs (5/8: first column, 6/8: second column). Boxes show interquartile ranges while the solid lines within each box represent the median. Whiskers cover up to 1.5 times the interquartile range while circles represent outliers beyond that.}
	\label{fig:quant_retro}
\end{figure}

	The quantitative evaluation on the whole test set shown in Figure \ref{fig:quant_retro} substantiates the findings from the qualitative comparisons. DRPF produces results that are significantly superior to POCS and zero-filling across both PFFs and $b$-values. For a PFF of 5/8 and averaged across both $b$-values, DRPF outperforms POCS by \mbox{4.64\,dB} and 0.0248 in terms of PSNR and SSIM, respectively. Since phase variations become more aggressive with stronger diffusion encoding, POCS tends to introduce more severe artifacts. Consequently, on images acquired at a $b$-value of \mbox{$800\,\text{s/mm}^2$}, POCS produces even lower SSIM scores than zero-filling for \mbox{$\text{PFF = 6/8}$} and outperforms zero-filling only slightly while having a larger variance for \mbox{$\text{PFF = 5/8}$}.
	
	\subsection{Prospectively sub-sampled liver DWI}
	\label{Res:2}
\begin{figure}[tbp]
	\centering
	\includegraphics[width=\textwidth]{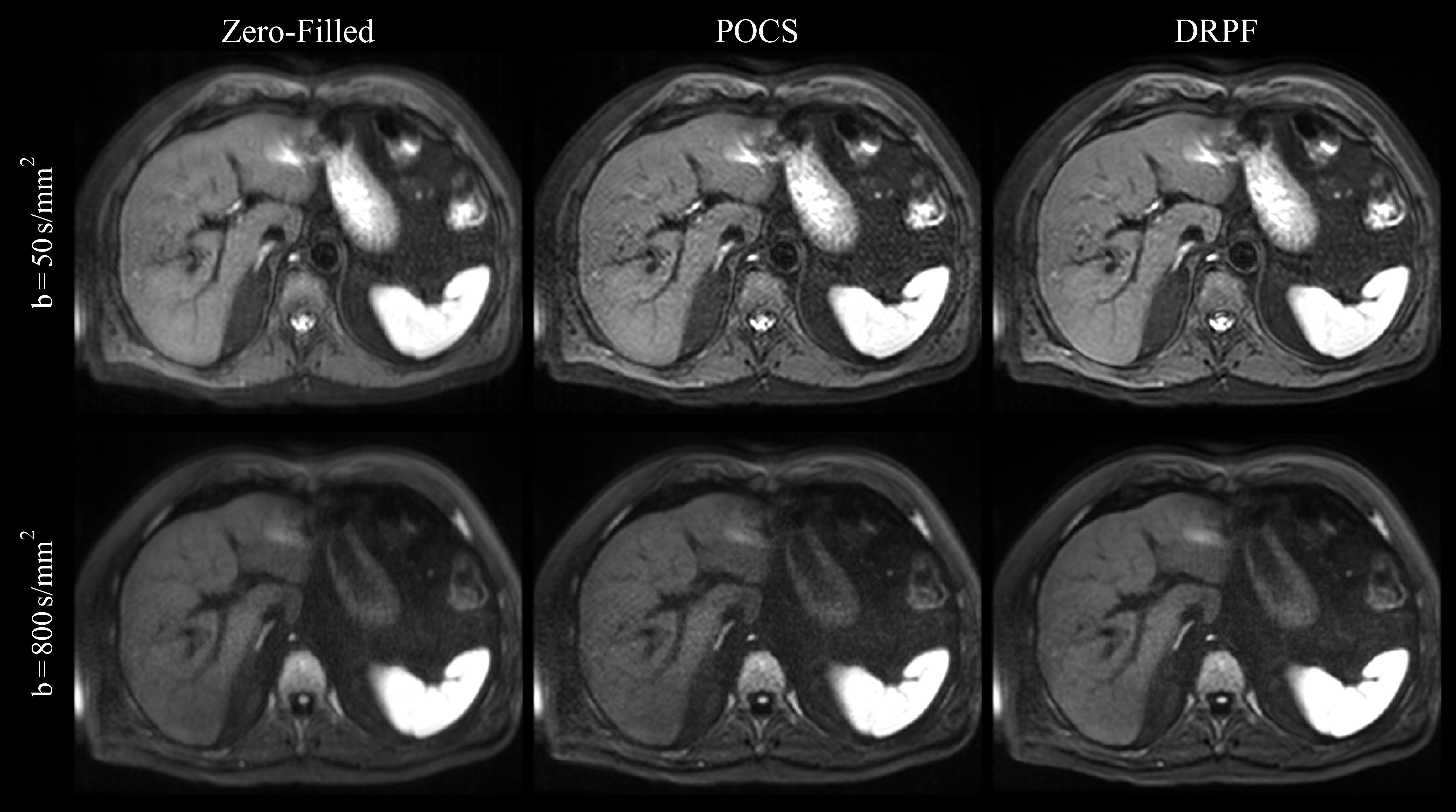}
	\caption{Qualitative evaluation of zero-filling, POCS and DRPF on a representative prospectively PF-sampled liver slice (\mbox{$\text{PFF} = 5/8$}, \mbox{1.5\,T}) acquired at $b$-values of \mbox{50\,s/mm$^2$} (top row) and \mbox{800\,s/mm$^2$} (bottom row).}
	\label{fig:qual_prospect}
\end{figure}

	As demonstrated in Figure \ref{fig:qual_prospect}, the proposed method also performs well on prospectively sub-sampled data. Compared to the zero-filled reconstruction, DRPF is able to restore resolution along the sub-sampled direction for both $b$-values. POCS reduces blurring as well but introduces additional noise and a mesh pattern. Representative reconstructions on prospective data of two additional subjects are presented in Supporting Information Figure \ref{sfig:qual_prospect_examples}.
	
\begin{figure}[tbp]
	\centering
		\includegraphics[width=\textwidth]{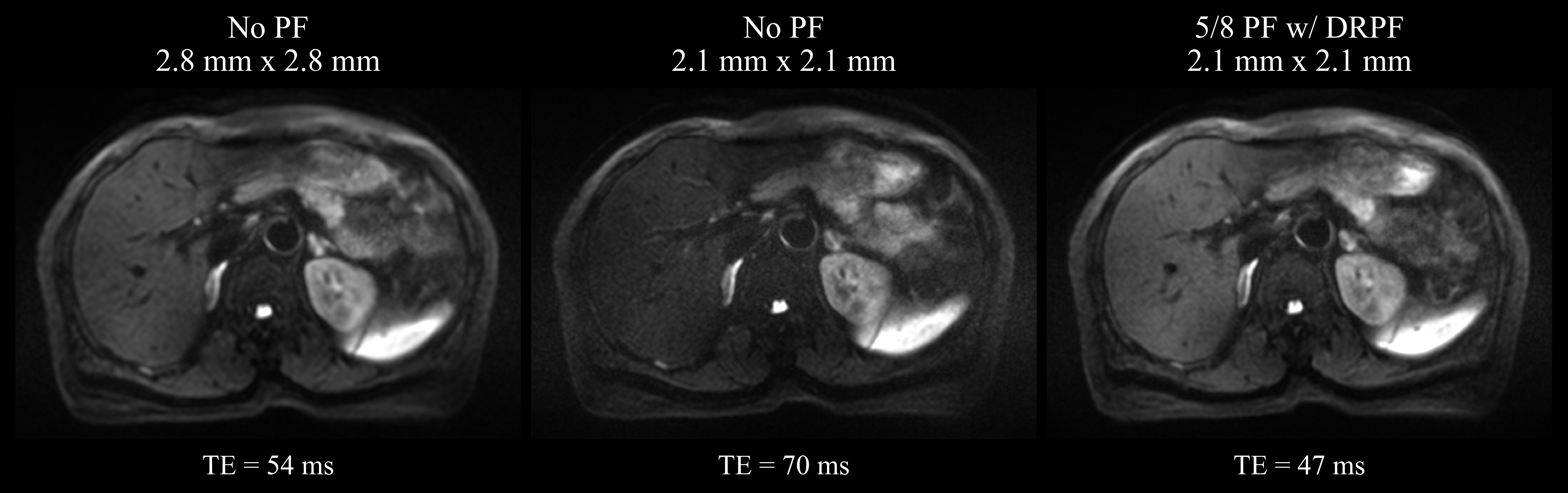}
	\caption{Qualitative comparison of three acquisitions at \mbox{3\,T} displayed with identical windowing: (left) standard resolution without PF-sampling, (middle) higher resolution without PF-sampling and (right) higher resolution with a PF-sampling of 5/8 reconstructed by the proposed method. Below each image the corresponding echo times are shown which were chosen as minimal as possible while leaving the TR constant across all three acquisitions.}
	\label{fig:qual_prospect_HR}
\end{figure}

	Figure \ref{fig:qual_prospect_HR} illustrates the effect of combining the proposed PF reconstruction with an acquisition of higher spatial resolution. Using a prospective PF-sampling of 5/8, TE can be reduced by almost one third. In fact, the resulting TE is even shorter than for a non-PF acquisition with the lower resolution. Consequently, by applying the proposed method to the sub-sampled data set, the reconstructed high-resolution image is comparable in SNR to the lower resolution image while not exhibiting visible sacrifices in terms of image sharpness compared to the high-resolution image acquired without PF.
	
	\subsection{Generalization to brain DWI}
\begin{figure}[tbp]
	\centering
		\includegraphics[width=\textwidth]{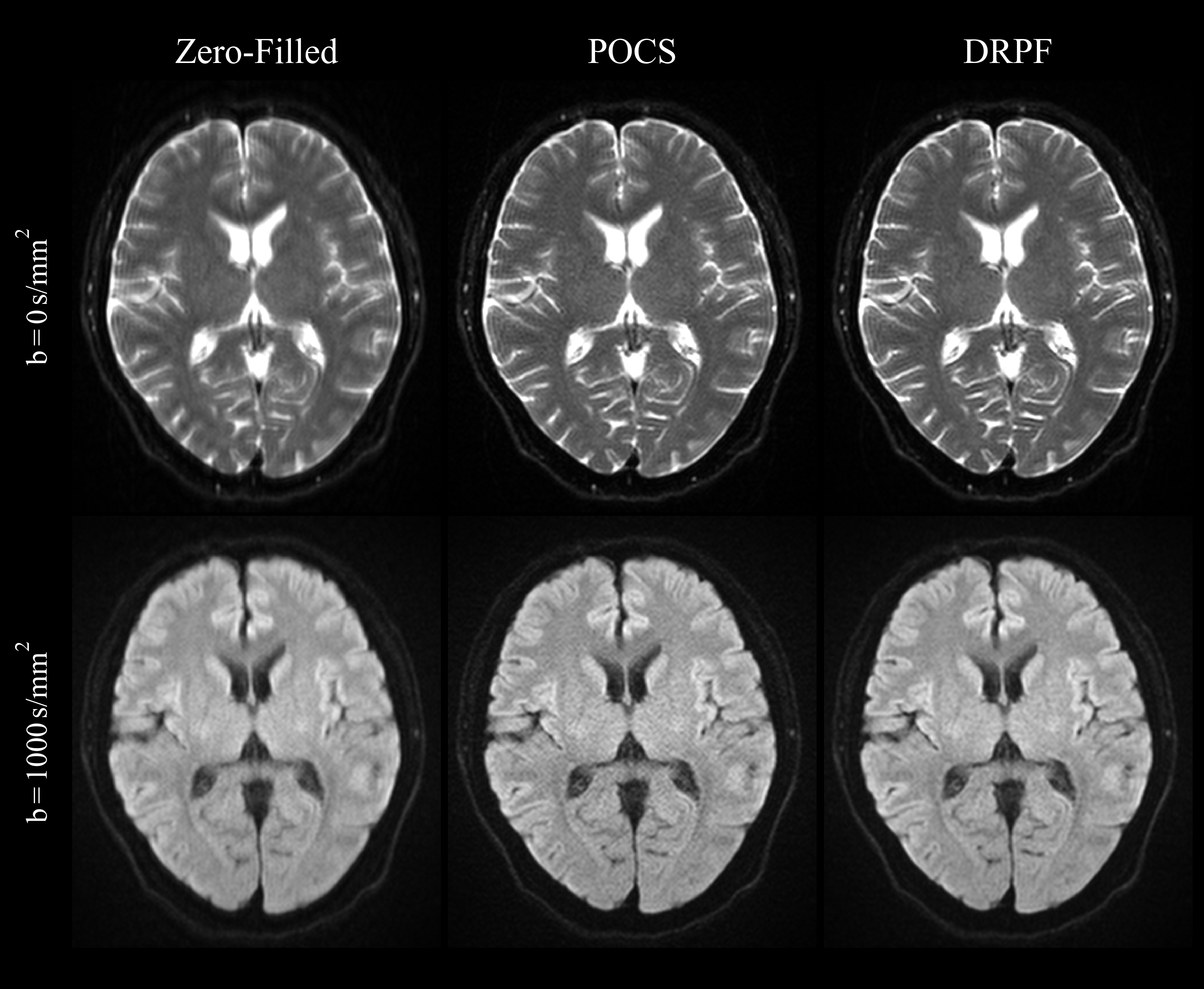}
	\caption{Qualitative evaluation of zero-filling, POCS and DRPF on a representative brain slice (\mbox{$\text{PFF} = 5/8$}, \mbox{3\,T}) acquired without (\mbox{$b=0$\,s/mm$^2$}, top row) and with (\mbox{$b=1000$\,s/mm$^2$}, bottom row) diffusion weighting.}
	\label{fig:brain}
\end{figure}

	Results on prospectively sub-sampled brain data are presented in Figure \ref{fig:brain}. Both POCS and DRPF alleviate blurring introduced by zero-filling effectively. However, POCS leads to noise amplification which becomes more profound for the higher $b$-value. Since phase variations are comparatively moderate in brain DWI, severe artifacts such as signal voids cannot be observed. Although being trained exclusively on liver data, DRPF is able to reconstruct brain images which feature sharp edges at tissue boundaries but at the same time appear more homogeneous in iso-intense regions compared to POCS. 
	
	\subsection{Benefits of joint reconstruction}
\begin{figure}[tbp]
	\centering
		\includegraphics[width=\textwidth]{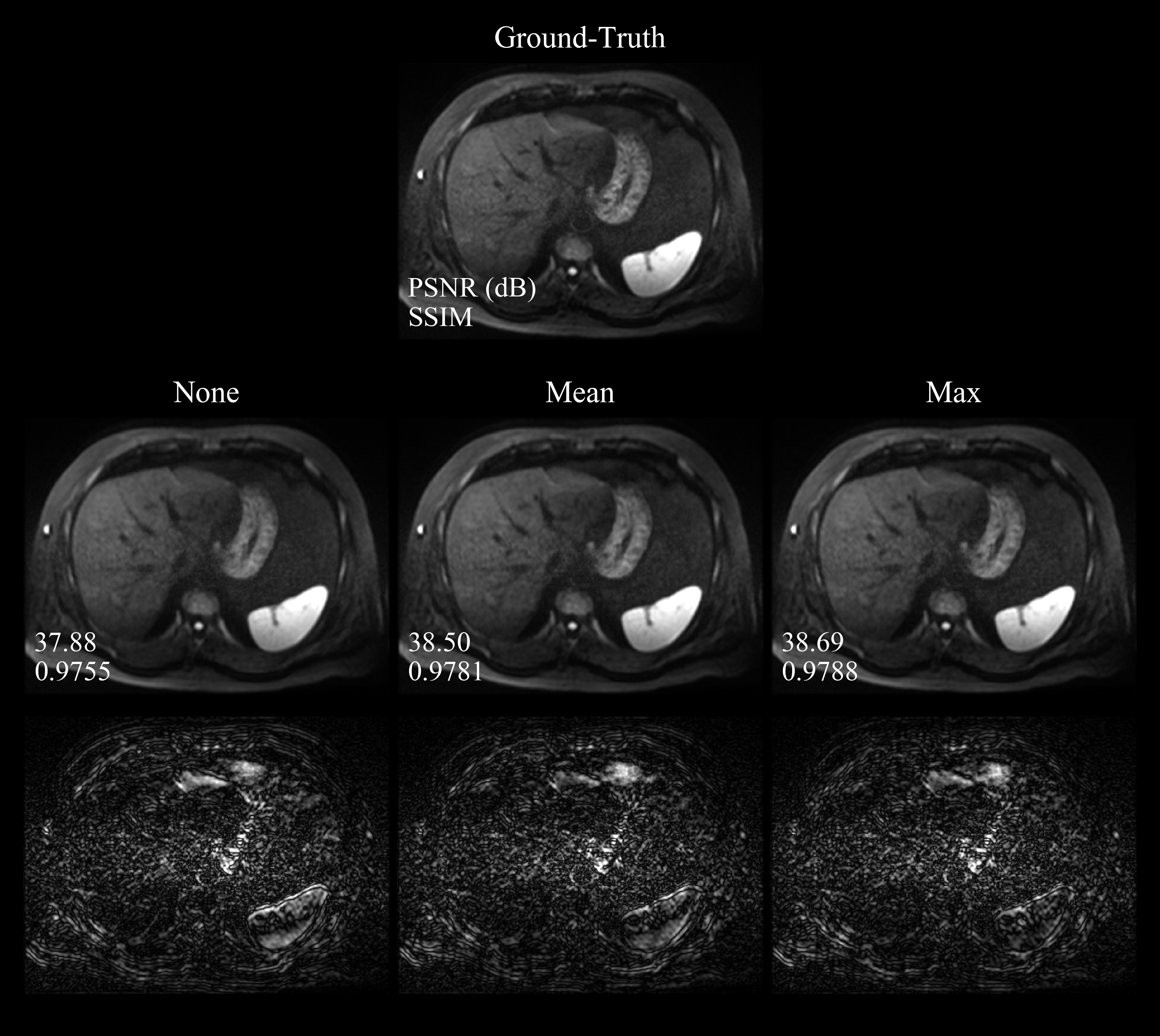}
	\caption{Reconstruction quality of different aggregation techniques on a representative retrospectively PF-sampled (\mbox{$\text{PFF} = 5/8$}) liver slice (magnitude average of 20 repetitions acquired at \mbox{3\,T} and a $b$-value of \mbox{800\,s/mm$^2$}). Top row: ground-truth image. Second and third row: reconstructions and corresponding difference images (magnified by a factor of 5) produced by network realizations which use no aggregation, mean and maximum aggregation, respectively.}
	\label{fig:comp_aggr}
\end{figure}

	As Figure \ref{fig:comp_aggr} implies, all three network realizations produce visually appealing reconstructions. However, error images reveal that the image obtained from the network which processes repetitions independently from each other exhibits a higher degree of residual blurring around tissue boundaries and some signal fluctuations (most apparent in the spleen). Using mean or maximum aggregation, both effects can be mitigated. While qualitative differences between the two aggregation approaches are minimal, maximum aggregation results in slightly improved PSNR and SSIM as confirmed by the quantitative evaluation on the whole test set provided in Table \ref{tab:comp_quant}. As the metrics imply, improvements of joint reconstruction are more substantial for data acquired at \mbox{$b=800$\,s/mm$^2$}. Given that a larger number of repetitions is available and individual repetitions are noisier, the greater benefit of joint reconstruction for higher $b$-values conforms to expectation.

\begin{table}[tbp]
	\centering
	\small
	\newcolumntype{C}[1]{>{\centering\arraybackslash}m{#1}}
	\newcolumntype{L}[1]{>{\raggedright\arraybackslash}m{#1}}
	\begin{tabular}{L{2.7cm}C{2.7cm}C{2.7cm}C{2.7cm}C{2.7cm}}\toprule
		& \multicolumn{2}{c}{$b =50$\,s/mm$^2$} & \multicolumn{2}{c}{$b = 800$\,s/mm$^2$}\\
		\cmidrule(lr){2-3}\cmidrule(lr){4-5}
		& PSNR (dB) & SSIM & PSNR (dB) & SSIM\\
		\midrule
		DRPF$_{\text{None}}$ & $39.12 \pm 3.48$ & $.9817 \pm .0142$ & $36.64 \pm 2.31$ & $.9725 \pm .0106$\\		
		DRPF$_{\text{Mean}}$ & $39.16 \pm 3.41$ & $.9820 \pm .0122$ & $37.44 \pm 2.36$ & $.9757 \pm .0107$\\		
		DRPF$_{\text{Max}}$ & $\mathbf{39.34 \pm 3.44}$ & $\mathbf{.9823 \pm .0122}$ & $\mathbf{37.57 \pm 2.33}$ & $\mathbf{.9762 \pm .0106}$\\
		\midrule
		Weight-sharing & $38.97 \pm 3.38$ & $.9815 \pm .0124$ & $37.40 \pm 2.35$ & $.9752 \pm .0108$\\
		Cascading & $39.12 \pm 3.41$ & $.9818 \pm .0123$ & $37.51 \pm 2.32$ & $.9756 \pm .0108$\\ 			
\bottomrule
	\end{tabular}
	\caption {PSNR and SSIM (mean $\pm$ standard deviation) on the retrospectively sub-sampled test set ($\text{PFF} = 5/8$) for different aggregation and unrolling techniques. Results are presented separately for the two employed $b$-values. Best results (\mbox{$p < 0.05$}) per metric and $b$-value are highlighted in bold. Note that both weight-sharing and cascading employ maximum aggregation.}
	\label{tab:comp_quant}
\end{table}

	\subsection{Comparison of unrolling strategies}
	As qualitative differences in the reconstructions produced by the different unrolling strategies are marginal, we focus on the quantitative comparison provided in Table \ref{tab:comp_quant}. The proposed unrolling method which uses recurrent convolutions consistently outperforms simple weight-sharing across $b$-values in terms of both SSIM and PSNR. The same applies with respect to cascading although it employs approximately five times the number of parameters of DRPF (2,227,530 vs.\ 474,450). However, the surplus in parameters still allows the cascaded network architecture to produce slightly superior results compared to the weight-shared network.

	\section{Discussion} \label{sec:discussion}
	Given that conventional PF techniques, such as POCS, rely on smoothness priors of the phase, they oftentimes introduce artifacts when applied to DWI which is prone to phase variations. Evaluations in this work showed effects like noise enhancement, the introduction of mesh patterns or even signal voids at strong PFFs (see Figures \ref{fig:qual_retro}, \ref{fig:qual_retro_single} and \ref{fig:qual_prospect}). All of those prohibit the use of PF-DWI in clinical practice or limit it to weak PFFs which do not cause significant reduction of TE in ssEPI sequences. The results of the conducted experiments imply that the proposed learning-based method achieves robust PF reconstruction of DW liver images with high-frequency phase structures. It was shown to outperform POCS and zero-filling quantitatively as well as qualitatively on a distinct testing set. In addition, the proposed method generalized well to brain DWI acquired at $b$-values that were not used in the training set. This implies that the network attempts to invert the blur kernel associated with PF-sampling rather than just learning liver anatomy.
	
	Further, it is noteworthy that recovery of high-frequency phase structures is possible although only magnitude information contributed to the objective function during training. This observation indicates that, similar to conventional methods, some form of phase regularization is exploited in order to reconstruct the missing data. However, in contrast to conventional methods, this is done in an implicit manner instead of using explicit smoothness constraints. We also hypothesize that the ability to recover signal voids arising from PF-sampling can be attributed to the restoration and utilization of high-frequency phase components as well as to the exploitation of information from other repetitions which potentially do not suffer from signal loss in the same area. In order to demonstrate that the network is not just filling signal voids empirically based on anatomy learned from training data, an experiment was performed in which an artificially generated signal void was placed into the repetition presented in Figure \ref{fig:qual_retro_single}. The corresponding results in Supporting Information Figure \ref{sfig:hole} show that the artificial signal `hole' in the liver parenchyma is deblurred but not filled, while the signal void in the left lobe resulting from PF-sampling is recovered as previously shown in Figure \ref{fig:qual_retro_single}.
	
	Further, the performance of the proposed method on prospectively sub-sampled data could be validated as well. Given the robust deblurring properties of DRPF, it is possible to omit almost one half of the $k$-space. This work demonstrated two ways of utilizing the associated TE reduction in combination with the proposed reconstruction technique. One was to simply minimize TE as much as possible in order to achieve DW images with high SNR without suffering from artifacts known from conventional methods (see Figure \ref{fig:qual_prospect}). Alternatively, it could be used to compensate for the TE increase associated with acquisitions of higher resolution (see Figure \ref{fig:qual_prospect_HR}). However, there are several other possible scenarios of utilizing the shortened echo-train. For example, the consequently increased SNR could be traded against scan time by reducing the number of acquired repetitions. Alternatively, the TE increase caused by bi-polar or motion-compensated diffusion preparation schemes could be countered by PF-sampling \cite{Karampinos, MODI}.

	The concept of \textit{Deep Sets} was employed in this work in order to provide the network with correlations within the set of available repetitions. Compared to a version of DRPF that did not exploit shared information, it could be shown that reconstruction quality could be improved by implementing aggregation of image features at certain points within the network. Here, maximum aggregation provided slightly superior results than mean aggregation. The fact that the incorporation of those operations is straightforward and does not increase the number of learned parameters renders this approach valuable in settings in which multiple realizations of an image are available.
	
	Since individual repetitions can be affected by imaging artifacts, the effects of those artifacts on the reconstruction of a selected repetition were investigated as well. In the context of DW ssEPI, signal dropouts due to cardiac pulsation are particularly prominent (see Supporting Information Figure \ref{sfig:pulsation_artifacts}). Using the proposed reconstruction pipeline with maximum aggregation across the set of image features (DRPF$_{\text{Max}}$), a specific repetition was reconstructed in two different configurations in which the remaining 19 repetitions of the set were either affected by signal dropouts (`corrupt') or not (`clean'), respectively. The results presented in Supporting Information Figure \ref{sfig:deepset} reveal that joint reconstruction with corrupt repetitions can diminish the image quality of the repetition at hand compared to reconstruction with clean repetitions only (\mbox{$-0.53$\,dB} PSNR, $-0.0066$ SSIM). However, individual reconstruction of repetitions (DRPF$_{\text{None}}$) can still be outperformed considerably (\mbox{$+0.56$\,dB} PSNR, $+0.0129$ SSIM), which substantiates the usefulness of the Deep Set concept regardless of the image quality of individual repetitions.

	Unrolled, model-based networks emerged as the state-of-the art architecture class for medical image reconstruction. As in our previous work \cite{DRPF}, the regularizer was implemented by a recurrent CNN here. Our experiments showed that the given architecture was able to quantitatively outperform two alternative unrolling strategies, namely weight-sharing and cascading. In addition, it does so with $K$ times less parameters than cascading, where $K$ is the number of unrolled iterations. Note that this is not necessarily an observation exclusive to PF reconstruction but could potentially be valid for other applications as well. However, in order to make a more general statement, further experiments are required which are beyond the scope of this work. One limitation of the comparison of unrolling strategies conducted in this work is the architectural difference between the recurrent unrolling approach and the other strategies. While the former uses ConvGRUs, the latter use a ResNet architecture. Since GRUs are recurrent by definition, a non-recurrent realization of it for comparison purposes would not be meaningful. Therefore, a ResNet architecture was chosen which was similar in the number of parameters.
	
	Using recurrent networks in unrolled architectures has been proposed in previous work as well. \textit{Recurrent Inference Machines} (RIMs) were introduced in \cite{RIM} and applied to accelerated MR reconstruction in \cite{RIM_MRI}. In contrast to this work which defines the iterative update rule via proximal splitting in a variational setting, RIMs are derived from a more Bayesian perspective in which the gradient is calculated according to a maximum-a-posteriori solution. Hence, in \cite{RIM} the recurrent CNN takes both the current estimate and its data-consistent transform as input instead of regarding regularization and data consistency as separate steps within an iteration. Further, a bi-directional recurrent network architecture was proposed in \cite{Qin} for unrolled reconstruction of dynamic cardiac MRI. Memory was propagated not only along iterations but also along the time dimension of the data. In contrast to this work which employed more sophisticated GRUs, basic Elman cells \cite{Elman} were used in \cite{Qin}.
	
	One general limitation of PF acquisitions is the increased risk of omitting the maximum $k$-space point which can be shifted away from the $k$-space center in data with large phase errors. Consequently, it would become impossible to fully recover the corresponding signal loss even for a Deep Learning-based solution. However, within our data set this scenario was very rare. While it did not occur at all for the lower $b$-value, the maximum $k$-space point was outside the region that would be sampled for $\text{PFF} = 5/8$ in only 0.53\,\% of the cases for $b=800$\,s/mm$^2$ (see Supporting Information Figure \ref{sfig:hist}). One of those cases is presented in Supporting Information Figure \ref{sfig:signal_loss} where the omitted $k$-space region carried the image information corresponding to the hyper-intense spleen which can be recovered only to a very limited degree by the proposed reconstruction method. Nevertheless, given that this problem only occurs in high $b$-values which are typically acquired with a relatively high number of repetitions, the effects on the averaged DW images are expected to be marginal as shown in Supporting Information Figure \ref{sfig:signal_loss_avg}. Here, the previously described signal loss in the spleen of the single repetition did not lead to perceivable intensity reduction in the corresponding area of the averaged image consisting of 20 repetitions. Despite the results shown in Supporting Information Figure \ref{sfig:hist}, local signal dropouts in PF-acquired data due to rigid bulk motion are still possible. In order to minimize the risk of that, respiratory-triggered or breath-held acquisitions can be employed instead of free-breathing acquisitions as performed in this work for the sake of faster data collection.
	
	As phase variations can be result of eddy currents which are typically stronger in prospectively sub-sampled data, clinical evaluation of the proposed method is planned involving a broader evaluation on prospective data by expert radiologists. As an outlook for future work, it could be also worthwhile to further investigate the aspects of recurrent network unrolling and joint reconstruction of image repetitions, for example, by employing more sophisticated aggregation techniques, such as attention-based methods \cite{Ilse, SetTransformer}. Also, the frequency of aggregations as well as their location within the network could be optimized for. Given its robustness with respect to phase variations, the presented method combined with potential re-training that penalizes phase errors explicitly could enable PF imaging in other applications in which phase is inherently non-smooth or even carries contrast information, such as in phase-contrast imaging.
	 
	\section{Conclusion} \label{sec:conclusion}
	This work demonstrated that robust PF reconstruction of coil-combined DW data is possible by means of an unrolled, model-based neural network. Using a regularization which is parametrized by recurrent convolutions and trained on retrospectively sub-sampled data, artifacts introduced by conventional PF methods due to ill-suited phase priors can be avoided. Reconstruction quality could be validated on retrospectively and prospectively sub-sampled data. Decent generalization to other anatomies and contrasts can be achieved which substantiates the ability of the method in approximating an inversion of the blurring operation. Accurate PF reconstruction combined with the associated TE reduction has the potential to enable acquisitions which might have been prohibited so far due to excessive echo-train lengths in ssEPI sequences and/or the lack of accurate PF reconstruction techniques.
	
	\section*{Conflict of Interest}
	
	Fasil Gadjimuradov receives PhD funding from Siemens Healthcare GmbH. Thomas Benkert and Marcel Dominik Nickel are employees of Siemens Healthcare GmbH.

	\clearpage
	\section*{List of Supporting Figures}
	\captionsetup[figure]{name=Supporting Information Figure}
	\setcounter{figure}{0}
	\renewcommand\thefigure{S\arabic{figure}}
		\begin{figure}[H]
			\centering
			\includegraphics[width=\textwidth]{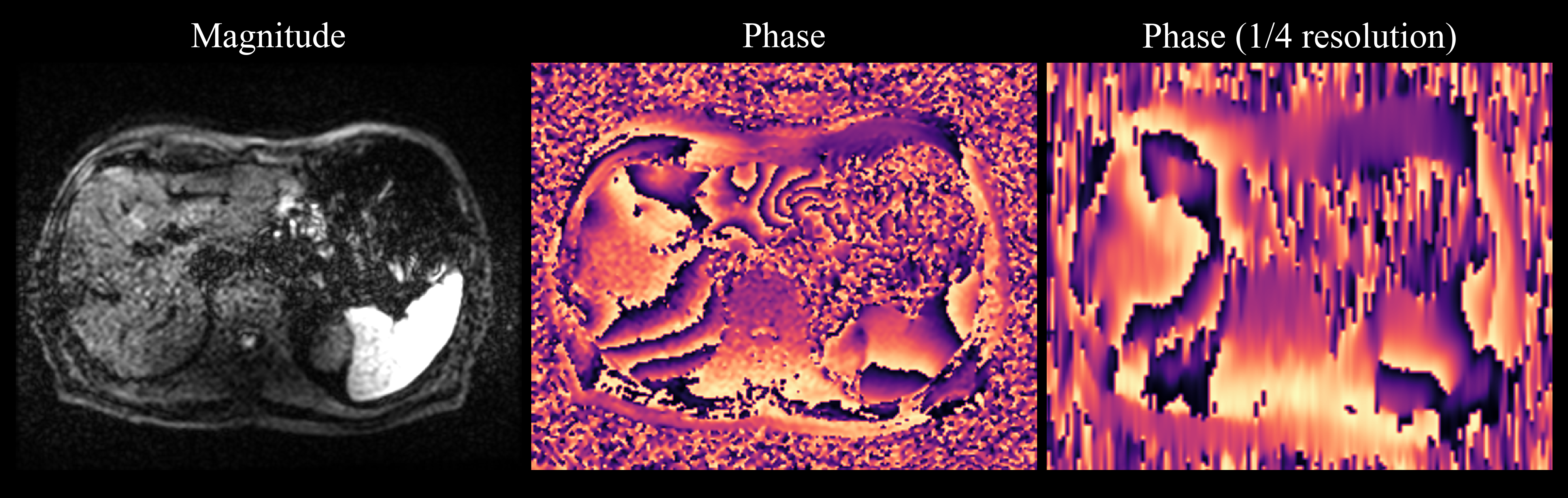}
			\caption{Phase variations in a representative repetition of a DW liver slice acquired at \mbox{3\,T} and a $b$-value of \mbox{800\,s/mm$^2$}: (left) magnitude image, (middle) phase image, (right) and the corresponding low-pass-filtered phase having only 1/4 of the original resolution. The latter represents a low-resolution phase estimate as it would be used in Homodyne or POCS for a PFF of 5/8. Note how high-frequency structures along AP, such as in the left liver lobe, are lost completely in the low-resolution phase.}
			\label{sfig:phase_liver}
		\end{figure}

		\begin{figure}[H]
			\centering
			\includegraphics[width=\textwidth]{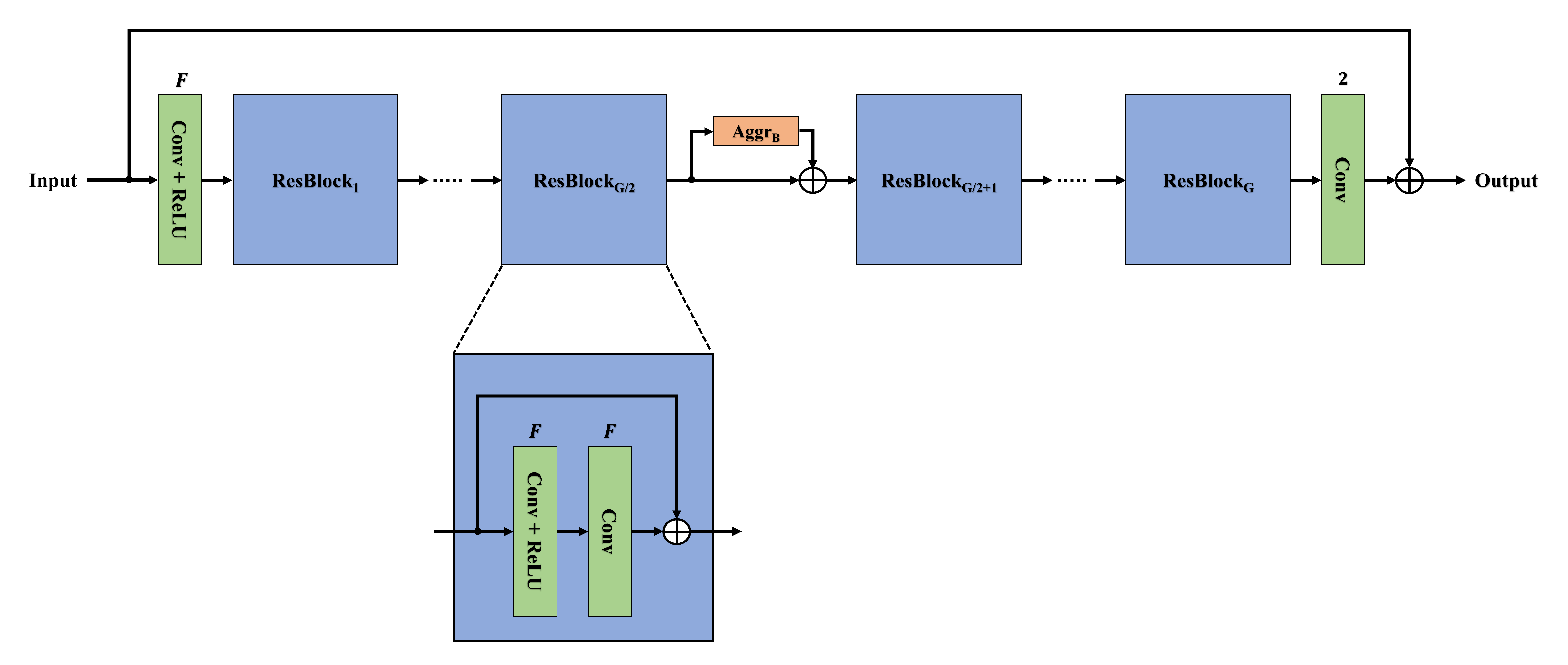}
			\caption{ResNet architecture used in the cascading and weight-sharing approaches. It consists of $G$ residual blocks (ResBlocks), each of which is composed of two convolutional layers (Conv) with $F$ feature channels. A batch aggregation (Aggr$_\text{B}$) is performed after half of the residual blocks. In order to match the number of parameters of the recurrent architecture, parameters were set as $G=6$ and $F=64$. Further, maximum aggregation along the batch of image features was employed.}
			\label{sfig:resnet}
		\end{figure}

		\begin{figure}[H]
			\centering
			\includegraphics[width=.9\textwidth]{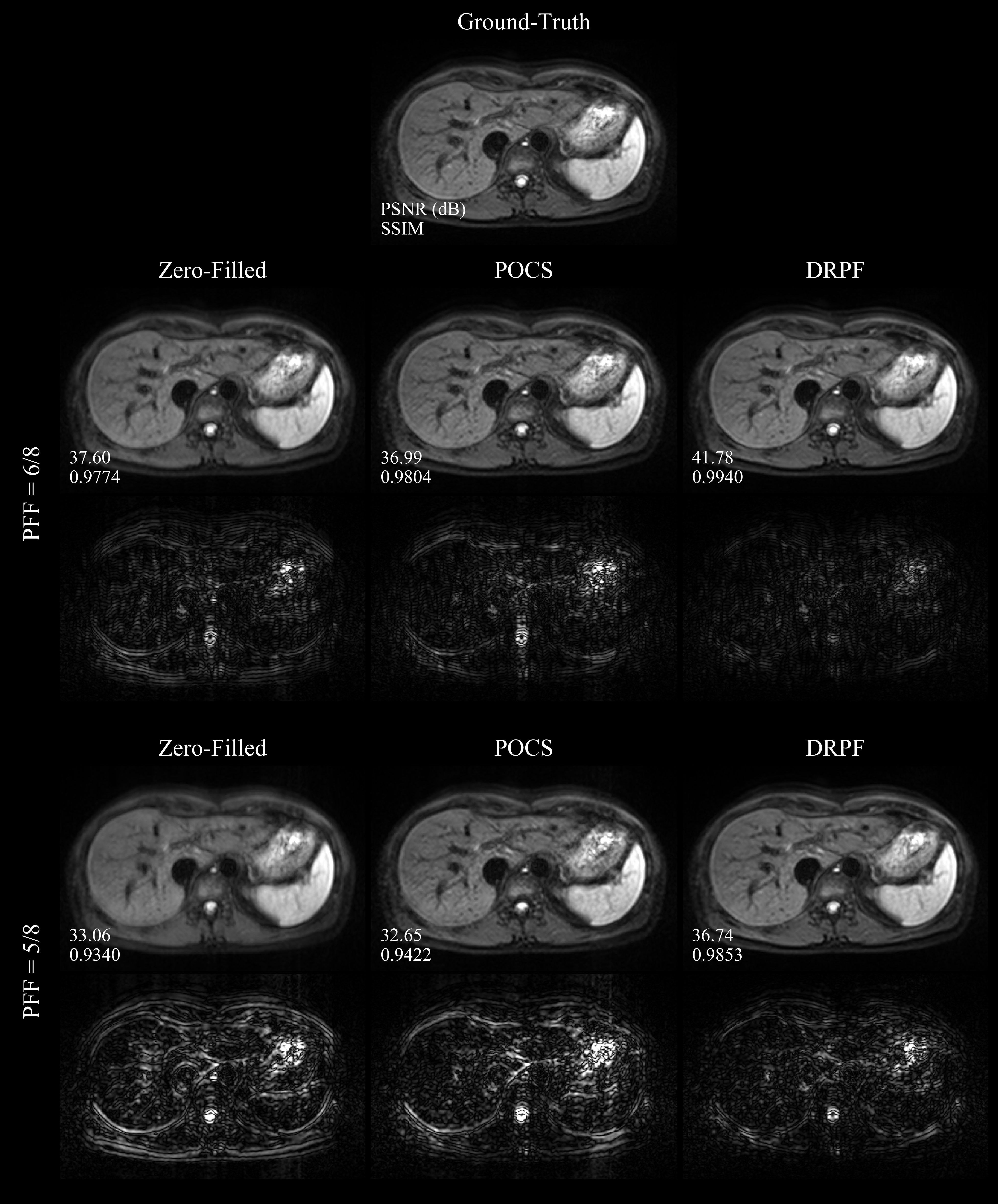}
			\caption{Reconstruction quality on retrospectively PF-sampled liver data (magnitude-average of 5 repetitions acquired at \mbox{1.5\,T} and a $b$-value of \mbox{50\,s/mm$^2$}). Top row: ground-truth image. Second and third row: reconstructions and corresponding difference images (magnified by a factor of 5) produced by zero-filling, POCS and DRPF at a PFF of 6/8. Fourth and fifth row: same as above for a PFF of 5/8.}
			\label{sfig:qual_retro_b50}
		\end{figure}
		
		\begin{figure}[H]
			\centering
			\includegraphics[width=\textwidth]{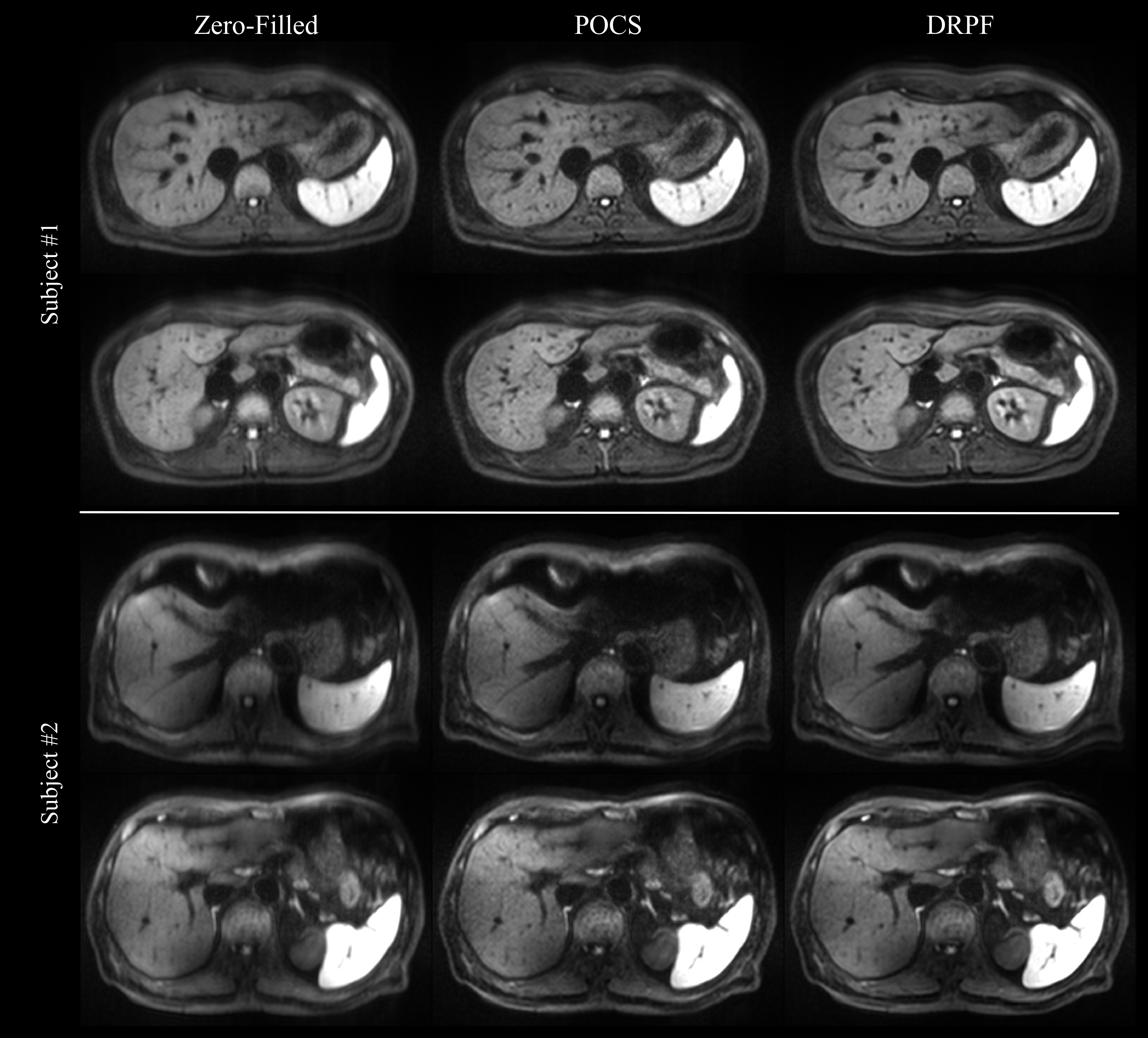}
			\caption{Reconstruction quality on representative prospectively PF-sampled liver slices (magnitude-average of 20 repetitions acquired at a $b$-value of \mbox{800\,s/mm$^2$}) from two different subjects. The upper and bottom rows show two slices per subject, respectively, reconstructed by zero-filling, POCS and DRPF.}
			\label{sfig:qual_prospect_examples}
		\end{figure}

		\begin{figure}[H]
			\centering
			\includegraphics[width=\textwidth]{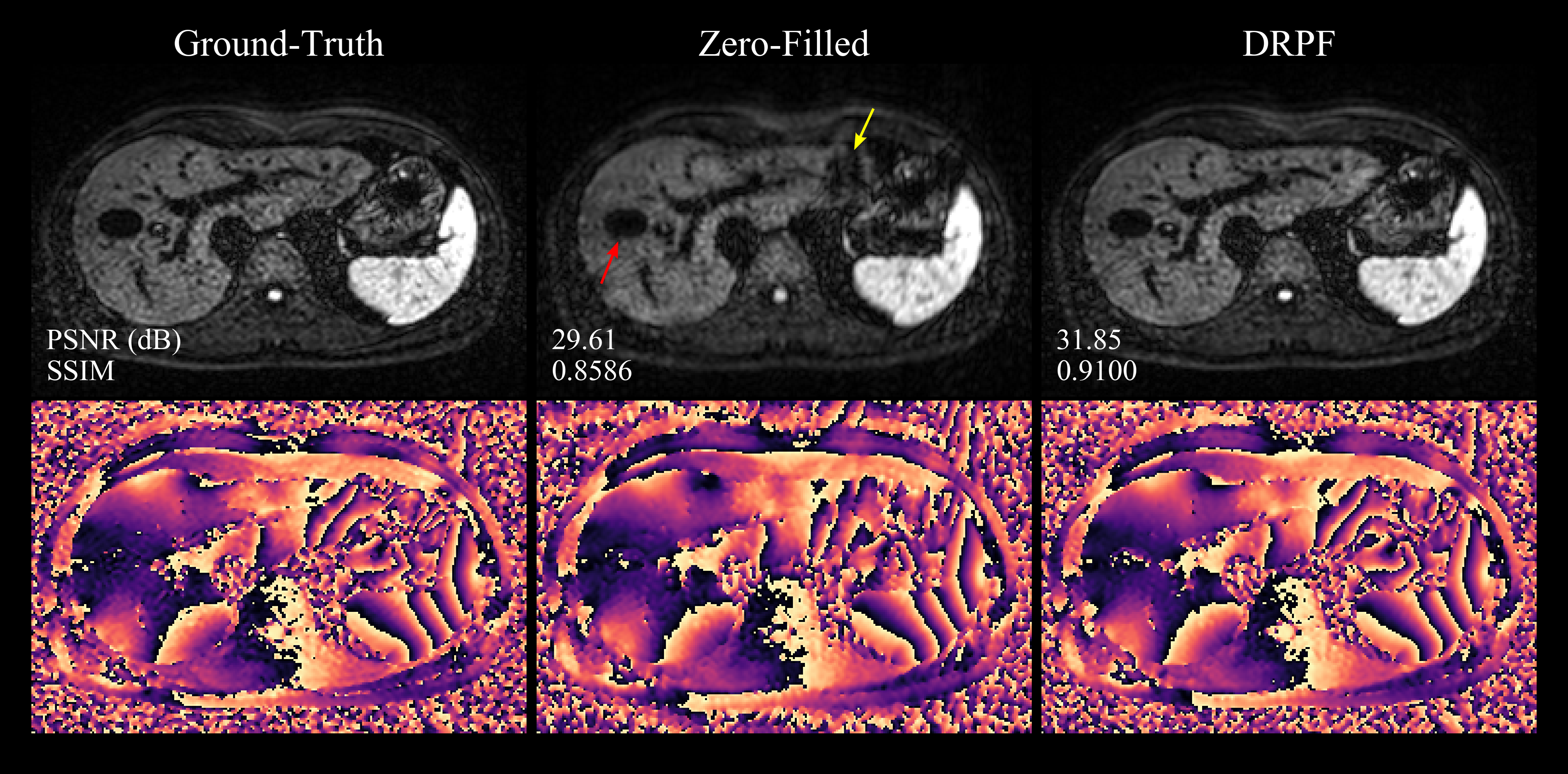}
			\caption{Reconstruction quality of DRPF on a retrospectively PF-sampled (\mbox{$\text{PFF} = 5/8$}) single repetition shown in Figure \ref{fig:qual_retro_single} with a simulated signal void within the liver parenchyma. First row: magnitude images. Second row: phase images. The signal void was generated by setting an elliptical region of the magnitude to zero and adding Rician noise consistent with the image noise. As a result, the zero-filled image exhibits two locations with signal voids within the liver: an artificial one (red arrow) and one due to PF-sampling (yellow arrow). While in the latter case, the respective signal can be recovered using DRPF, the signal remains close to zero in the former.}
			\label{sfig:hole}
		\end{figure}

		\begin{figure}[H]
			\centering
			\includegraphics[width=\textwidth]{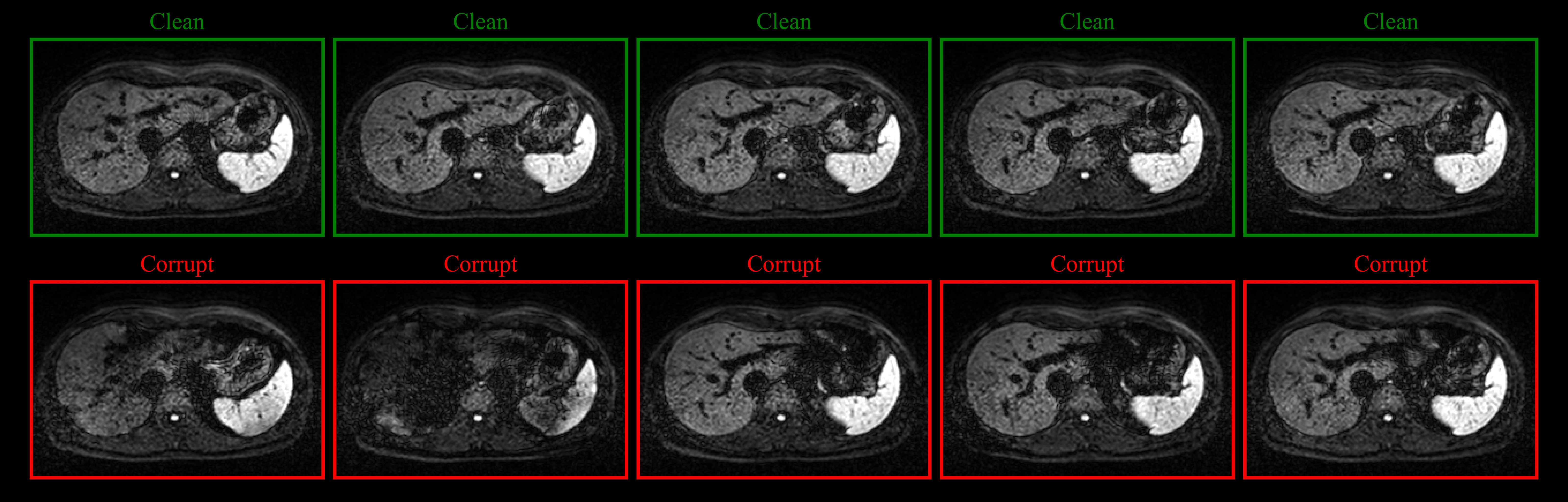}
			\caption{Ten representative repetitions of a liver slice acquired at \mbox{1.5\,T} and a $b$-value of \mbox{800\,s/mm$^2$}. The five repetitions in the bottom row are affected by motion-induced signal dropouts of different severity. Note that these dropouts appear in the fully-sampled images. Also note that the repetitions exhibit good positional alignment despite a free-breathing acquisition because only a small sub-set of the total available repetitions (60) is shown. Only repetitions from a similar motion state were selected such that the differences between the repetitions mainly stem from pulsation-related signal dropouts.}
			\label{sfig:pulsation_artifacts}
		\end{figure}
		
		\begin{figure}[H]
			\centering
			\includegraphics[width=\textwidth]{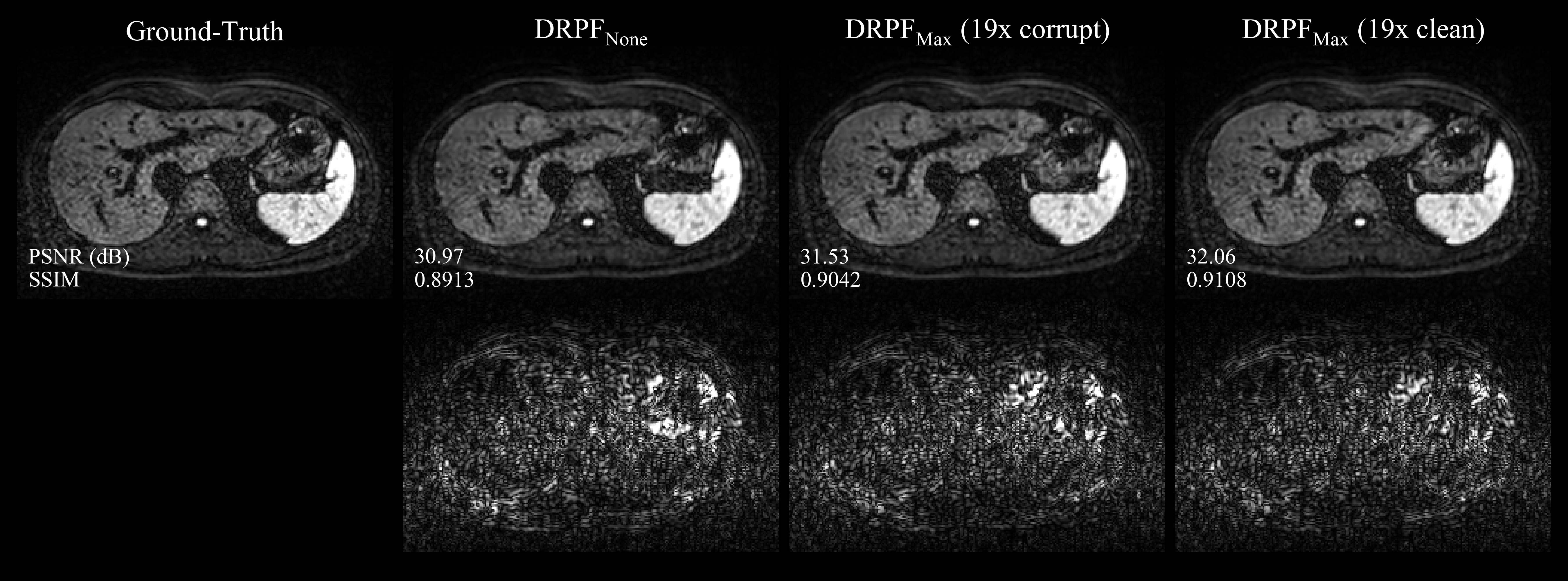}
			\caption{Reconstruction quality of different aggregation configurations on a retrospectively PF-sampled (\mbox{$\text{PFF} = 5/8$}) single repetition shown in Figure \ref{fig:qual_retro_single}. Three configurations were compared: separate reconstruction of repetitions (DRPF$_\text{None}$) and joint reconstruction (DRPF$_\text{Max}$) where the 19 remaining repetitions of the set were either affected by motion-induced signal dropouts (`corrupt') or not (`clean'). The first and second row show the magnitude images and the corresponding difference images with respect to the ground-truth (magnified by a factor of 5), respectively.}
			\label{sfig:deepset}
		\end{figure}
			
		\begin{figure}[H]
			\centering
			\includegraphics[width=\textwidth]{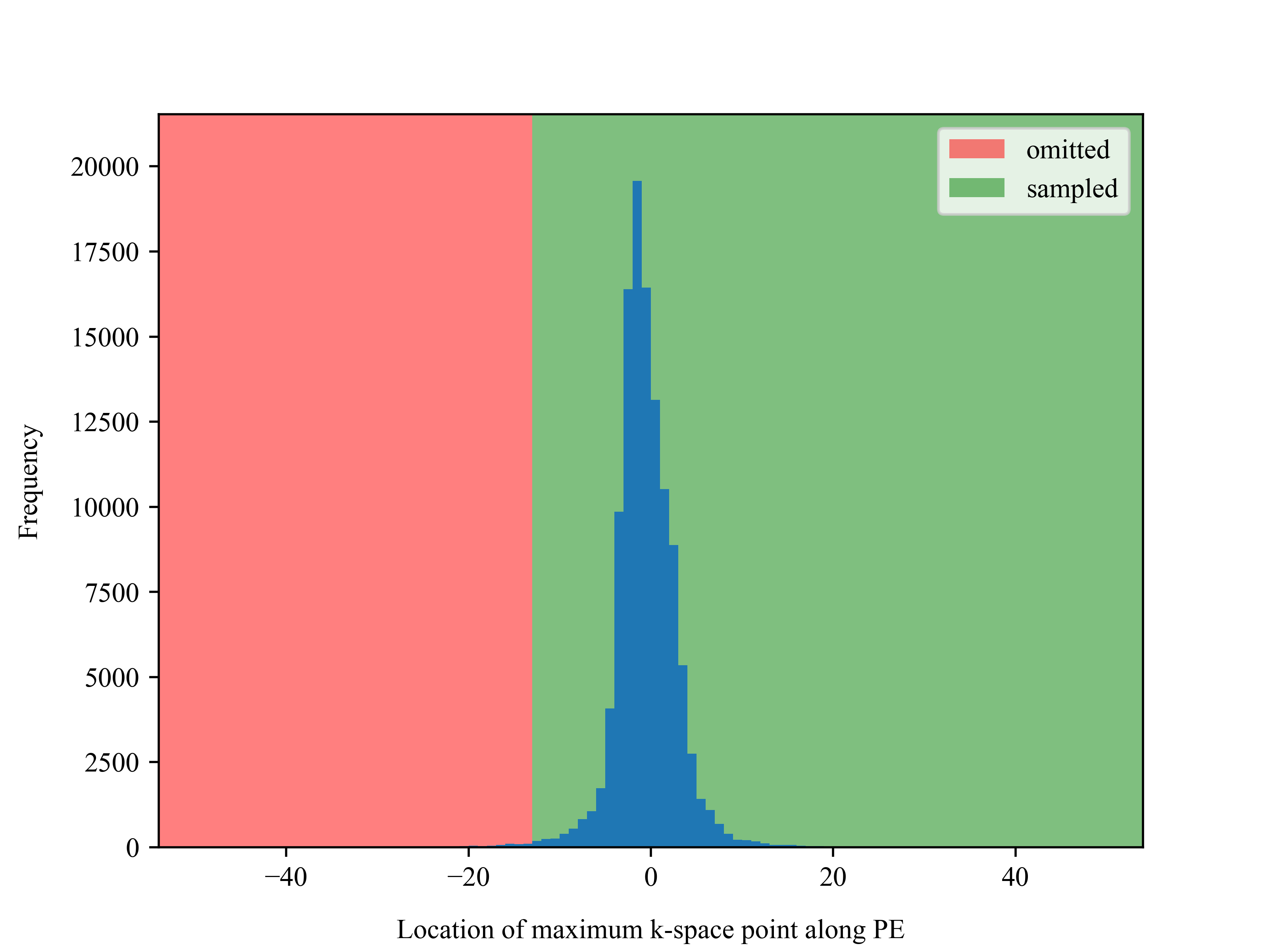}
			\caption{Histogram of maximum absolute frequency locations along PE in all available individual repetitions acquired at \mbox{$b = 800$\,s/mm$^2$}. In only 700 out of 133,200 repetitions (0.53\,\%), the maximum frequency location was outside of the region that would have been sampled for a PFF of 5/8 and 108 PE steps.}
			\label{sfig:hist}
		\end{figure}

		\begin{figure}[H]
			\centering
			\includegraphics[width=\textwidth]{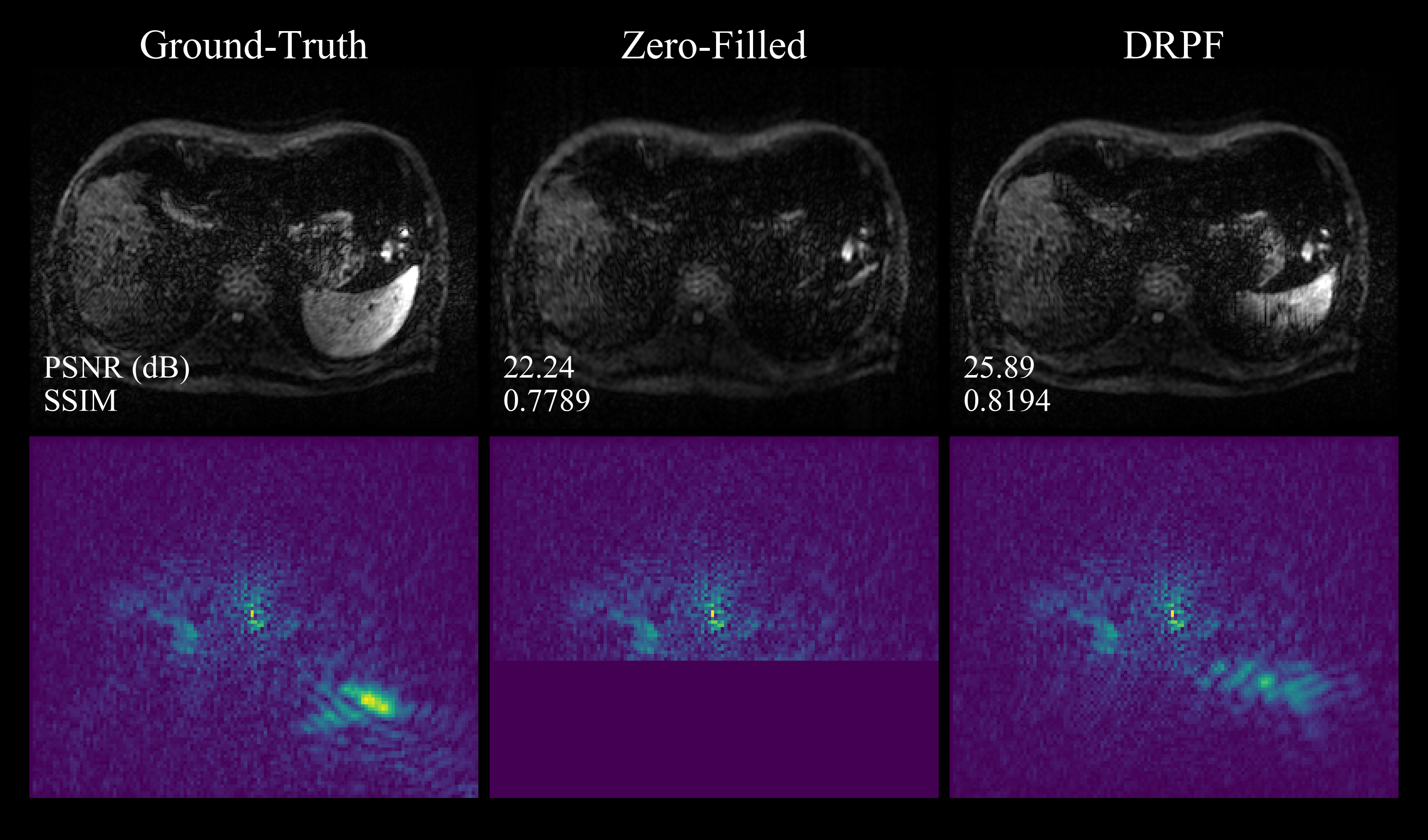}
			\caption{Reconstruction quality of DRPF on a retrospectively PF-sampled repetition (\mbox{$\text{PFF} = 5/8$}, \mbox{$b = 800$\,s/mm$^2$}) in which the maximum $k$-space point corresponding to the signal of the hyper-intense spleen is not covered due to sub-sampling. First row: magnitude images. Second row: magnitude of $k$-spaces.}
			\label{sfig:signal_loss}
		\end{figure}
		
		\begin{figure}[H]
			\centering
			\includegraphics[width=\textwidth]{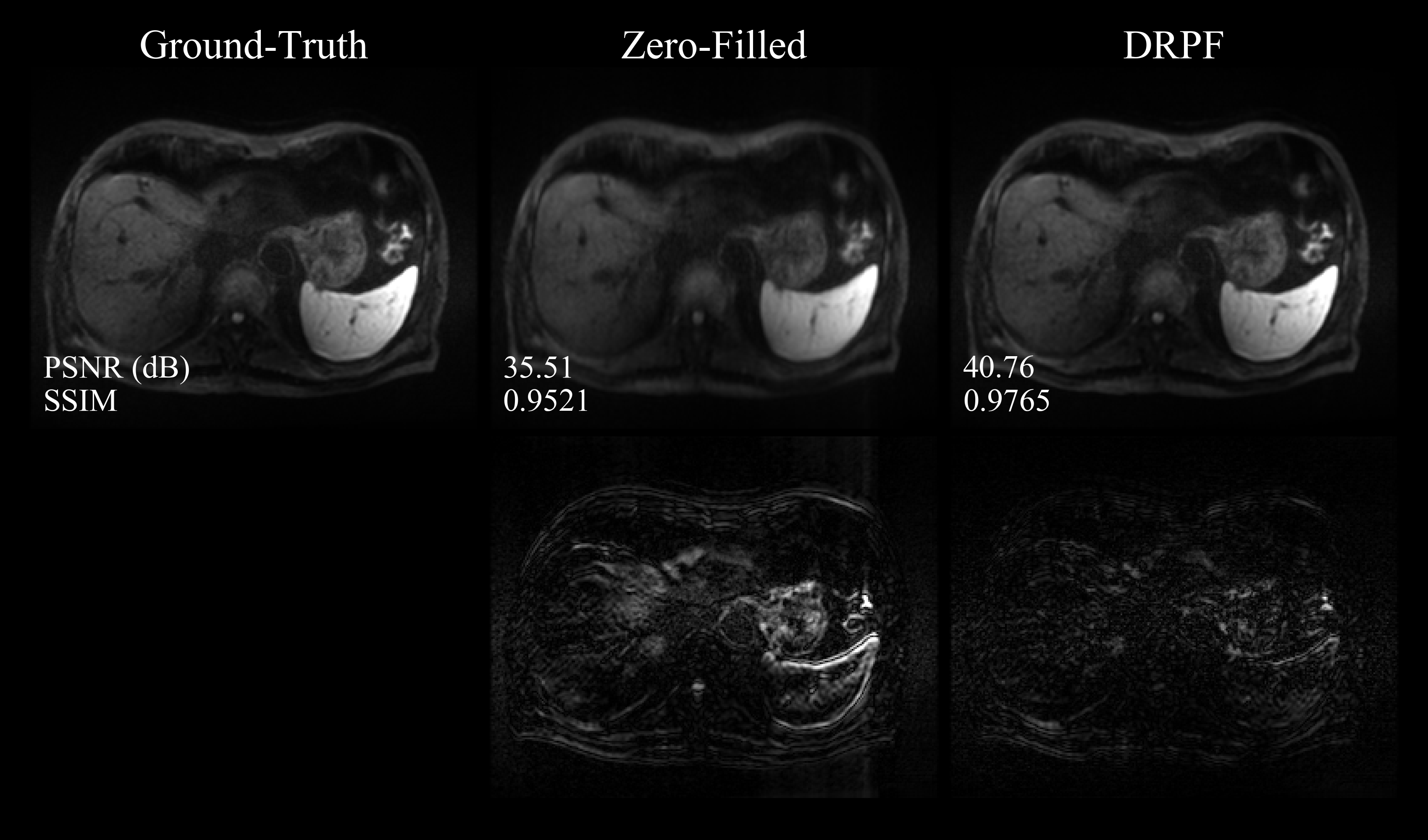}
			\caption{Reconstruction quality of DRPF on a retrospectively PF-sampled liver slice (\mbox{$\text{PFF} = 5/8$}, \mbox{$b = 800$\,s/mm$^2$}) consisting of 20 repetitions, one of which is shown in Supporting Information Figure \ref{sfig:signal_loss}. The first and second row show the magnitude images and the corresponding difference images with respect to the ground-truth (magnified by a factor of 5), respectively.}
			\label{sfig:signal_loss_avg}
		\end{figure}
	
	\clearpage
	\section*{List of Supporting Tables}	
	\captionsetup[table]{name=Supporting Information Table}
	\setcounter{table}{0}
	\renewcommand\thetable{S\arabic{table}}    

	 	\begin{table}[H]
			\centering
			\small
			\newcolumntype{C}[1]{>{\centering\arraybackslash}m{#1}}
			\newcolumntype{L}[1]{>{\raggedright\arraybackslash}m{#1}}
			\begin{tabular}{L{3.5cm}C{3cm}C{3cm}C{3cm}C{3cm}}\toprule
				& LR w/o PF & HR w/o PF & HR w/ 5/8 PF\\\midrule
				FOV (mm$^2$) & $380 \times 308$ & $380 \times 308$ & $380 \times 308$\\	
				Matrix size & $134 \times 108$ & $180 \times 146$ & $180 \times 146$\\
				Resolution (mm$^2$) & $2.8 \times 2.8$ & $2.1 \times 2.1$ & $2.1 \times 2.1$\\
				TR (ms) & 6,300 & 6,300 & 6,300\\
				TE (ms) & 54 & 70 & 47\\
				\bottomrule
			\end{tabular}
			\caption{Comparison of relevant parameters of the DWI acquisitions described in \ref{sec:prospect_data} collecting data of low (LR) and high resolution (HR) with and without prospective PF-sampling.}
			\label{stab:acqparam_prospect}
		\end{table}
		
		\begin{table}[H]
			\centering
			\small
			\newcolumntype{C}[1]{>{\centering\arraybackslash}m{#1}}
			\newcolumntype{L}[1]{>{\raggedright\arraybackslash}m{#1}}
			\begin{tabular}{L{4.5cm}C{3cm}}
				\toprule
				$b$-values (s/mm$^2$) & $[0, 1000]$\\
				\# of diffusion directions & [1, 4]\\
				\# of repetitions & [3, 4]\\
				FOV (mm$^2$) & $230 \times 230$\\	
				Matrix size & $192 \times 192$\\
				Resolution (mm$^2$) & $1.2 \times 1.2$\\
				Slice thickness (mm) & 5\\
				B0 (T) & 1.5\\
				TR (ms) & 6,700\\
				TE (ms) & 81\\
				Parallel acceleration & 2\\
				PF factor & 5/8\\
				\bottomrule
			\end{tabular}
			\caption {Acquisition parameters of the DW brain data set.}
			\label{stab:acqparam_brain}
		\end{table}
	
\end{document}